\documentclass[11pt]{article}
\usepackage{amsmath,amssymb}
\usepackage{threeparttable}
\usepackage{etoolbox}
\usepackage{enumerate}
\usepackage{multirow}
\usepackage{wrapfig}
\usepackage{titlesec}
\usepackage{float}
\usepackage{indentfirst}
\usepackage{url} 
\usepackage{algorithmic}
\usepackage[colorlinks,citecolor=blue,urlcolor=blue]{hyperref}
\usepackage[utf8]{inputenc}
\usepackage[table]{xcolor}
\usepackage{caption}
\usepackage{xpatch}

\usepackage{graphicx}
\graphicspath{ {/Users/selena/Desktop/dissertation/res/real data/new data} }

\usepackage{amsthm}
\DeclareUnicodeCharacter{0301}{\'{e}}
\usepackage{psfrag,epsf}
\usepackage{enumerate}
\usepackage{mathtools}
\newtheorem{lemma}{Lemma}

\usepackage{tocloft}

\newlength{\mylen}

\renewcommand{\cftfigpresnum}{\figurename\enspace}
\renewcommand{\cftfigaftersnum}{:}
\usepackage{booktabs}
\usepackage{indentfirst}
\usepackage{url}
\usepackage[colorlinks,citecolor=blue,urlcolor=blue]{hyperref}
\allowdisplaybreaks

\usepackage[boxruled,scleft]{algorithm2e}

\RestyleAlgo{boxruled}
\usepackage[T1]{fontenc}
\usepackage{mathtools} 

\numberwithin{equation}{section}

\usepackage{setspace}

\allowdisplaybreaks
\usepackage[
backend=biber,
style=apa,uniquename = false,
]{biblatex}
\title{A bibLaTeX example}
\usepackage{titlesec}
\usepackage{hyperref}

\begin{document}
\def\spacingset#1{\renewcommand{\baselinestretch}%
{#1}\small\normalsize} \spacingset{1}

\spacingset{1.4}

\begin{center}
   \textbf{\Large{\\The co-varying ties between networks and item responses via latent variables.}}
\end{center}

\vspace{0.2in}

\begin{center}

Selena Wang\footnote{Corresponding Author, selewang@iu.edu},
    Department of Biostatistics and Health Data Science, Indiana University School of Medicine\\
    Plamena Powla, Department of Biostatistics and Health Data Science, Indiana University School of Medicine\\
    Tracy Morrison Sweet, Department of Human Development and Quantitative Methodology, University of Maryland \\
    Subhadeep Paul, 
    Department of Statistics,  
    The Ohio State University

\end{center}

\begin{abstract}
Relationships among teachers are known to influence their teaching-related perceptions. We study whether and how teachers' advising relationships (networks) are related to their perceptions of satisfaction, students, and influence over educational policies, recorded as their responses to a questionnaire (item responses). We propose a novel joint model of network and item responses (JNIRM) with correlated latent variables to understand these co-varying ties. This methodology allows the analyst to test and interpret the dependence between a network and item responses. Using JNIRM, we discover that teachers' advising relationships contribute to their perceptions of satisfaction and students more often than their perceptions of influence over educational policies. In addition, we observe that the complementarity principle applies in certain schools, where teachers tend to seek advice from those who are different from them. JNIRM shows superior parameter estimation and model fit over separately modeling the network and item responses with latent variable models. 
\end{abstract}

\section{Introduction}


Relationships among teachers are known to influence educational outcomes. Teachers' relationships with each other are positively correlated with teaching performances, perceptions of trust for peers and students' achievements, among others. In particular, \textit{teaching networks}, composed of relationships in which teachers discuss their instructional practice with peers are vital to school climates. Teachers are more willing to engage with school policies if they feel more connected with other teachers and principals \parencite{moolenaar2010occupying}. They are more willing to change and adopt new teaching practices if they are part of the teaching network and if the new practice has been adopted by teachers in their network \parencite{penuel2013organization}. On the other hand, school climates are known to influence the teaching network. Teachers tend to have larger networks following professional development programs and tend to be more connected in schools that previously engaged in school-wide initiatives \parencite{weinbaum2008going}. 

Network analysis is a collection of quantitative methods modeling relationships (edges) among entities (nodes). Network analysis has far-reaching applications in nearly all domains of science, including studies of World-Wide Web, power grid systems, trade and economic relationships, human health, brain networks, metabolic and gene expression networks, ecological networks, and social networks, among others \parencite{ zhang2019scale,janssen2006toward, baggio2016multiplex,wuchty2014controllability,jeong2000large, bastiampillai2013depression, cohen2000resilience, wang2022resilience,jackson2008social,epskamp2017generalized,bullmore09,rubinov10,mcpherson2001birds,paul2022causal}.  

In this article, we study the relationship between teachers' networks and school climates. We study whether and how teachers' advising interactions are related to their perceptions of schools. The network includes information about who the teachers seek advice from during the school year. The school climates perceived by the teachers are measured as their responses to a set of survey questions called item responses. Among the survey questions, three categories of questions were given to all teachers: perceived satisfaction with the school, perceived quality of the students, and the perceived impact on school policies. To test and understand the relationship between the network and the perceived school climate, we propose a novel methodology that jointly analyzes a network and item responses from a questionnaire. Possible questions that can be answered with our methodology include whether the advice-seeking network or the latent dimensions of the advice-seeking network are related to the item responses or the latent dimensions of the item responses. 
Our methodology is applicable for the joint analysis of any data consisting of a network and item responses.

While several studies have looked into the relationship between the network of teachers and the school climates, rarely do they assess the relationship using fully specified statistical models, (e.g., see \textcite{broda2021using}). More often, summary statistics such as node degrees (the total number of advice-seeking relationships for each teacher) are used as inputs for further assessment ( \textcite{bonsignore2011power}). A downside to this type of analysis is that two drastically different networks can be reduced to similar summary statistics, and thus the structures of the networks beyond the summary statistics are subsequently lost. For example, networks with similar node degree distributions can have different community structures \parencite{paul2022null}. Focusing on node degrees implies that important network information can be lost. 

The proposed methodology falls under a broad class of fully specified statistical models that aim to analyze the relationship between a network of individuals and their attributes. We use the term attribute loosely to represent information about characteristics, property, quality, and behaviors of the nodes, etc. One way to achieve this goal is to incorporate the attributes of nodes as covariates in a network model, for example in exponential random graph models
(ERGMs) \parencite{lusher2013exponential}, stochastic blockmodels \parencite{mele2019spectral,sweet2015incorporating} and latent space models \parencite{hoff2002latent}. In these network models with covariates, social networks are the dependent variables, and the attributes are included in the model as predictors. Alternatively, we can estimate the conditional\footnote{The effects are estimated conditional on the observed network.} effects of the networks on the attributes by regressing the attributes on the networks. The latter are called the social influence models, where attributes are the dependent variables and information from the networks is included in the model as predictors \parencite{sweet2020latent}. 

A different approach is to develop a joint modeling framework for network data and attributes. \textcite{palestro2018tutorial} distinguishes a joint modeling framework from a ``separate model'' or a ``submodel'' by stating that the goal of a joint framework is to model the covariation of the parameters of separate models in order to capture the important information in each data. Although their discussions about the joint framework are rooted in the neural and behavioral research, we believe that their conceptual framework is helpful for understanding the joint models of a network and attributes as well.  \textcite{palestro2018tutorial} describes three types of joint models: Integrative, Directed, and Covariance. In the Integrative joint framework, two types of data are explained together using a shared set of parameters.  In the Directed joint framework, the parameters describing one type of data are modulated by the parameters describing the other type of data, and the direction of the modulation is known. In the Covariance joint framework, parameters describing one type of data are related to parameters describing the other type through a shared distribution.

In this article, we develop a Covariance joint framework for a network and attributes. In \textcite{wang2023joint}, an Integrative joint framework, the joint latent space model (JLSM) was proposed for a network and attributes. The authors used shared latent variables to describe potential dependence, or the lack of independence, between the network and attributes.  The JLSM imposes a strong assumption about the data generation process---the latent variables corresponding to the person nodes in the network are also used for the item response data. Our reason for not choosing the Directional joint framework is that it requires prior knowledge about the direction of the dependence between the network and attributes, while in practice, the relationship is often correlational. We believe that a Covariance joint framework is the most flexible framework.

In the proposed covariance joint framework, we assume two sets of latent variables for individuals, with one set responsible for the network formation and the other responsible for item responses. The latent variables describing the network formation (also called latent network dimensions) and the latent variables describing the attribute behavior data share a multivariate normal distribution with a joint covariance matrix. This approach allows the information in the attributes to inform the estimation of the network's latent variables and vice versa. The degree of the association reflected in the covariance matrix of the multivariate normal distribution is the strength of the dependence between the network and attributes.

The proposed method is different from that in \textcite{fosdick2015testing}, which models the correlation between latent variables estimated from a network and simple attribute data. Meanwhile, item responses or responses to questionnaires or test items are complicated types of attribute data. A questionnaire item is intentionally constructed by people, often researchers (although AI-generated items recently became available), to yield a response from a respondent, which can serve as a measurement for some psychological construct such as ability, traits, characteristics, behaviors, etc. These item responses should be modeled as indicators of latent psychological constructs. Not providing a latent variable theory about the generation of the item responses oversimplifies the problem and is a lost opportunity to model the measurement process. Thus, in such scenarios with multidimensional and dependent item responses, the method proposed in \textcite{fosdick2015testing} is not applicable.

The proposed joint model allows for simultaneous inference about the dependence between and within the network and attributes of the persons, a significant improvement over the network models with covariates or the social influence models. These models only allow inferences about the relationship in one direction. Explaining the social network based on the item responses does not allow us to make inferences about the dependence among the item responses, and explaining the item responses based on the network does not allow us to make inferences about the dependence in the network. To make inferences about the relationships at the latent level on both sides, we should use a joint framework. Another drawback of the network model with covariates and of the social influence model is that at least one side, either the network or the attributes is assumed to be fully observed. There may be interest in predicting the network relations or attributes of a node when both its network information and attributes are partially missing. We can also use the joint modeling framework to provide predictions when there are missing observations in both the network and attributes.

Modeling the dependence between the network and item responses using correlated latent variables with a joint multivariate normal distribution has its advantages. If there is dependence between the network and attributes, and each modeling component is reasonably valid, we can expect an improvement in parameter estimation using the joint model compared with using the separate models. This is because estimation of each component influences and informs the other. The joint framework is also an improvement over existing approaches where the relationship between the network and attributes is quantified in two steps, e.g., first extracting the latent factor and then estimating the relationship as the correlation between the latent factor and attributes or using the latent factor as the predictor of a regression model in the second step. With the exception of \textcite{paul2022causal}, the uncertainty associated with estimating the latent factors in the first step is typically ignored when estimating the relationship in the second step. Lastly, modeling dependence as correlations is a flexible and intuitive option. No prior knowledge about the direction of the dependence is needed.

To reiterate, we propose a novel joint statistical model for the network data and item responses (JNIRM---joint network and item response model). In the following sections, we first describe the motivating application, then the model. We propose a MCMC algorithm to estimate the proposed model, and then we report results for the applications. Lastly, we discuss and conclude.

\section{Motivating Application}

In this section, we present a real-life example, where the research question is whether and how a network and an item response matrix are related. The data set was collected in 2015 from teachers in $14$ schools in Nebraska \parencite{pustejovsky2009question,pitts2009using}. It includes the social network data from teachers and their responses to a school climate questionnaire. Among the teachers were $14$ principles, $14$ school administrators with teaching responsibilities and $378$ with mostly teaching responsibilities. Among the teachers, $87.93 \%$ were females. The racial composition is as follows: $98.52\%$ of the sample were Caucasian, $1.23\%$ were Hispanic, and less than $1\%$ were Mexican, African American, and Native American.  These teachers were involved in students' learning from pre-kindergarten to the $6$-th grade and are engaged in a variety of subjects, including Art, Dance/Drama, English, History, Physical Education, Music, Science, Math, Social Studies with an average of $10.87$ years of teaching experience.

\subsection{Advice-seeking Network}
\begin{figure}
\begin{center}
    \includegraphics[width=.8\textwidth]{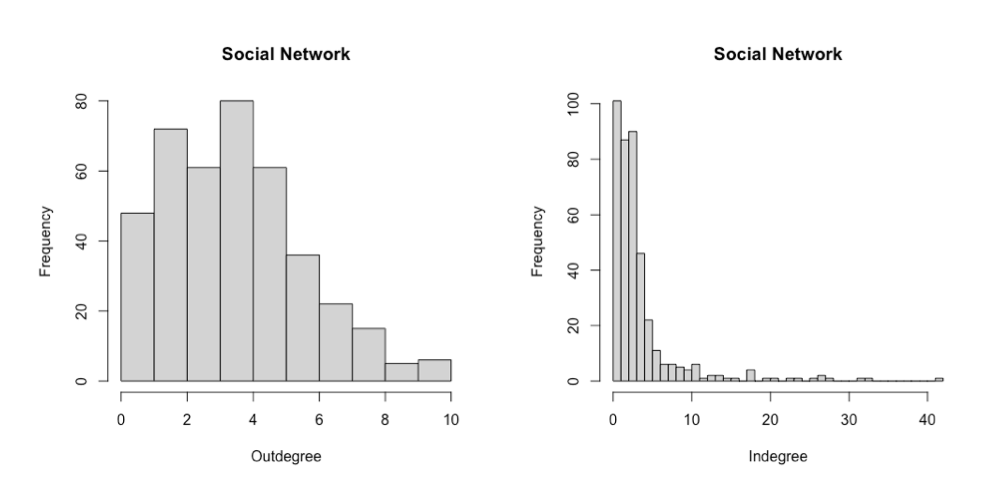}
  \end{center}
  \vspace{-25pt}
      \caption{Degree distribution for the advice-seeking network. Frequencies of number of persons one seeks advice from are outdegree, and frequencies of number of persons seeking advice from the same person are indegree. }
\label{degreepic}
\end{figure}

Each teacher was asked to indicate to whom she or he turned to for advice during the school year, and at most $12$ advisors were reported per teacher though many listed less than $12$. For each advisor, the teacher was asked to describe the type of advice sought and the frequency of the advice-seeking behaviors as (1) Daily, (2) Weekly, (3) Monthly, or (4) A few times per year. In this project, we present the advice-seeking network as a binary adjacency matrix $\boldsymbol{X}$. The presence of an advice-seeking relationship, regardless of the frequency, is counted as an observed edge. In the adjacency matrix, $x_{a,b}=1$ if teacher $a$ seeks advice from teacher $b$. 
The advice-seeking relationships are directional, and $\boldsymbol{X}$ is asymmetric. Teacher $a$ seeking advice from teacher $b$ does not imply that teacher $b$ seeks advice from teacher $a$. The degree distributions are shown in Figure \ref{degreepic}. The indegree distribution is more skewed than the outdegree distribution suggesting that most of the advice was sought from a few experienced advisors, and most teachers were not asked for advice.

\begin{figure}
\begin{center}
    \includegraphics[width=.5\textwidth]{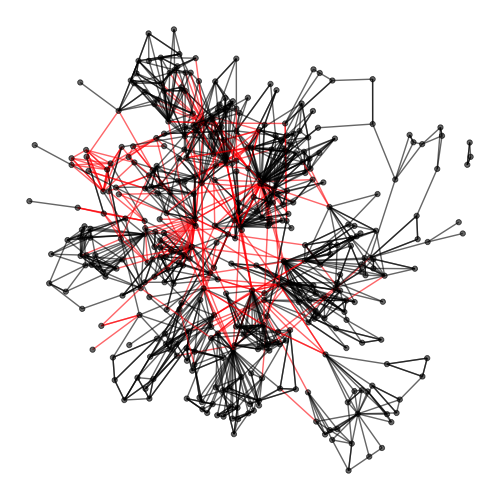}
  \end{center}
  \vspace{-25pt}
      \caption{An advice-seeking social network among teachers. Nodes are teachers, and an edge is present between two teachers from the same school (black) and different schools (red) when one seeks advice from the other. The figure does not show the direction of the relationship.}
\label{advice2}
\end{figure}

In total, $79.39 \%$ of the advice-seeking relationships were between teachers in the same school, $16.63\%$ were between teachers from different schools, $1.5\%$ were from professional development providers, $1.5\%$ were from community members, and less than $1\%$ of the advice were from friends or family members. In our study, we include advice-seeking relationships among those who were teachers in these $14$ schools and who were also asked about their perceptions of the school climates; we excluded teachers without perceptions of the school climate. In Figure \ref{advice2}, edges are colored black when the advice-seeking relationships are between teachers of the same school, and edges are colored red when the relationships are between teachers from different schools.


\subsection{School Climate Questionnaire}

The teachers were presented with the School Staff Social Network Questionnaire (SSSNQ), which includes a variety of items about the school climate. We chose $16$ items on the questionnaire that were distributed to all teachers; of the $16$ items, $4$ items are about the teacher’s overall satisfaction with the school (we will call these items the satisfaction items), $7$ items concern the teacher’s perceived influence over school policies (we will call these items the policy items), and $5$ items concern the teacher’s perceptions of the students (we will call these items the student items). Each of the three subscales of the climate questionnaire includes items designed to measure a different construct: the satisfaction, the perceived influence over school policies, and the perception of students. We describe the complete list of questions under the three different constructs in the supplementary materials.

The satisfaction items and the policy items were measured on a five-point Likert scale, and the student items were measured on a four-point Likert scale. For the five-point scale, the response options were ``Strongly Disagree", ``Disagree", ``Neutral", ``Agree", and ``Strongly Agree"; for the four-point scale, the response options were ``None", ``A little", ``Some'' and ``A great deal''. The Likert scale data were treated as continuous using a linear model. We chose not to dichotomize the data because we found that after dichotomization, the data were not sufficiently informative for the latent structure of the items. Future research should extend the joint model for the network and the item responses with the graded response model or the generalized partial credit model \parencite{samejima2016graded,muraki1990fitting}.

\subsection{Dimensionality of the School Climate Questionnaire}\label{rotation}

We explored the dimensions of the item responses with a principle component analysis (PCA) of the correlation matrix. We applied the PCA to the data from all schools. Although a multigroup factor model or a multilevel factor model may be more appropriate, we considered a PCA as sufficient for an exploratory analysis, given that the inter-school differences were rather small (see the supplementary materials for a comparison of 4 schools). The eigen values showed that there might be $3$ or $4$ components. The eigen values were $4.811$ $(28\%), 2.052$  $(14\%), 1.521$ $(9\%)$, and $1.060$ $(7\%)$ for the first four components, respectively.

 \setlength{\arrayrulewidth}{.3mm}
\setlength{\tabcolsep}{36pt}
\renewcommand{\arraystretch}{1.2}
\begin{table}[!ht]\caption{Rotated Loadings of the PCA}
 \begin{center}
\begin{tabular}{lccc} 
\toprule
&\multicolumn{3}{c}{ Loadings}\\
 \cmidrule(lr){2-4}
Item & Student& Policy & Satisfaction \\     
 \midrule
1 & \textbf{0.667}	&0.049	&0.091\\
2 & \textbf{0.731}&	-0.003&	0.146\\
3 & \textbf{0.291}	&-0.051	&-0.062\\
4 & \textbf{0.563}	&0.076	&0.045\\
5& \textbf{0.767}	&-0.071&	-0.158\\
6& -0.054	&\textbf{0.591}&	0.159\\
7& 0.056	&\textbf{0.621}	&0.014\\
8&-0.018	&\textbf{0.881}	&-0.107\\
9 &-0.045	&\textbf{0.866}	&-0.155\\
10 & 0.032&	\textbf{0.736}&	-0.107\\
11& 0.036&	\textbf{0.556}&	0.164\\
12 & -0.007&	\textbf{0.630}&0.188\\
13 &0.084&	0.127&	\textbf{0.671}\\
14&-0.132	&0.034	&\textbf{0.866}\\
15 &-0.085	&-0.031	&\textbf{0.922}\\
16& 0.183&	0.001&	\textbf{0.672}\\
\bottomrule
\end{tabular}
\begin{tablenotes}
      \linespread{1}\small
          \item Note: The loadings that are expected to be high based on the target matrix are in bold.
    \end{tablenotes}
\label{pcaloadings}
\end{center}
\end{table} 

Given that there are three subscales in the questionnaire, we rotated the first three components to a theoretical target structure that corresponds to the three subscales (with a zero loading on the components that do not correspond to the subscale an item belongs to). All items of the same subscale were given the same high loadings in the target theoretical structure (i.e., a value of 2 –- different high values yield the same rotation result). The rotation was performed using Michael Browne’s target rotation, which is also implemented in CEFA \parencite{browne2008cefa}, using the target.rot function in the psych R package \parencite{revelle2015package}. 

 \setlength{\arrayrulewidth}{.3mm}
\setlength{\tabcolsep}{10pt}
\renewcommand{\arraystretch}{1.2}
\begin{table}[!ht]\caption{Congruence Coefficients}
 \begin{center}
\begin{tabular}{lccc} 
\toprule
Coefficients & Student& Policy & Satisfaction   \\     
 \midrule
 Student & 0.944  &0.000        &0.017\\
 Policy &0.000 & 0.980    &    0.035\\
 Satisfaction& 0.017  &0.036    &    0.953\\
\bottomrule
\end{tabular}
\label{cong}
\end{center}
\end{table}

The loadings of the rotated components are shown in Table \ref{pcaloadings}. The loadings that are expected to be high based on the theoretical structure are in bold. The rotated components showed a clear correspondence with the theoretical structure, with the items from the same subscale loading on the same component and having low loadings on the other two components. To assess the similarity between the rotated components and the theoretical structure, we also calculated the congruence coefficients (see the results in Table \ref{cong}). The congruence coefficients are the cosines between the rotated components and the components defined by the theoretical structure. Table \ref{cong} shows that the congruence coefficients are high between the corresponding components and the theoretical structure. Therefore, we believe that the three rotated components correspond to the three subscales. We will apply the same target rotation to the results from the joint analysis of the network data and questionnaire data in the application section. The joint model will be described next.

\section{The Joint Network and Item Response Model}

In this section, we describe the joint network and item response model (JNIRM). Conceptually, the model is a joint model for two kinds of data, and it consists of three components: (1) a model for the network data, (2) a model for the item response data, and (3) a joint component that connects the previous two components. We will explain each of the three components.

\subsection{Modeling the Network Component}

In a latent space model, a node's information is summarized by low-dimensional continuous latent variables and can sometimes be visually represented in a low-dimensional space. The latent variable value for a node can be written in vector form with the number of elements in the vector equaling the number of dimensions. We restrict the scope of our discussion to vector latent space models that use the vector product to model the connections between nodes (see \textcite{wang2021recent} for details about the vector latent space models). One reason for using vector latent space models is to maintain consistency with the item response component of the joint framework. Models for item responses often use the vector product between a latent person variable and an item slope in the item response function, e.g., the two-parameter IRT model and the factor models \parencite{de2004framework}.

We can summarize the relational information of person $a$ as a sender in the network using a $K$-dimensional vector $\boldsymbol{u}_a$, called the latent position of sender $a$. Similarly, we can summarize the relational information of person $b$ as a receiver in the network using a $K$-dimensional vector $\boldsymbol{v}_b$, called the latent position of receiver $b$. The latent position for a node is the coordinate of the node on the corresponding latent network dimension. In this article,  we model the probability of a connection from a sender $a$ to a receiver $b$ using the vector product $\boldsymbol{u}_a^T\boldsymbol{v}_b$, called the multiplicative UV effect. In network analysis, other forms of the vector products also exist, e.g., the bilinear effect \parencite{hoff2005bilinear}. The multiplicative UV effects model allows for separate sender and receiver \textit{role specific latent variables} for a node, and is therefore more suitable for a directed network. For a directed network with a low level of reciprocity, the multiplicative UV effect is a better choice than the bilinear effect. We use the multiplicative UV effect to model the directed relationships in the advice-seeking network.

\noindent \textbf{Continuous edge weights.} We introduce the vector latent space model for a network with continuous quantitative edge weights. Let $\boldsymbol{X}$, be an $N \times N$ adjacency matrix representing the network, where $N$ is the number of persons in the network. The edge value $X_{a,b}$ represents the relationship from sender $a$ to receiver $b$, $a,b =1,...,N$ and $b \neq a$. 
For a directed network with continuous edge weights, the latent space model can be written as: 
\begin{align}
&X_{a,b} = \delta +\boldsymbol{u}_a^T \boldsymbol{v}_b + e_{a,b},  \qquad \begin{psmallmatrix} \boldsymbol{u}_{a} \\\boldsymbol{v}_a\end{psmallmatrix} \overset{iid}{\sim} N(\boldsymbol{0},\boldsymbol{\Lambda}_{u}),   \nonumber \\
& \{ (e_{a,b}, e_{b,a}): a < b \}   \overset{iid}{\sim}  N_2 \left( \boldsymbol{0}, \sigma^2_e \begin{pmatrix} 
 1 &\rho\\
 \rho & 1
\end{pmatrix}
\right),  \label{n1}
\end{align}where $\delta$ is a fixed intercept parameter accounting for the density of the network. As mentioned before, $\boldsymbol{u}_a$ and $\boldsymbol{v}_b$ are latent positions for sender $a$ and receiver $b$. The error for the edge from sender $a$ to receiver $b$ is $e_{a,b}$, and its covariance structure with $e_{b,a}$ follows Equation \ref{n1}. We call $\sigma^2_e$ the error variance and $\rho$ the within-dyad\footnote{A pair of nodes.} correlation. The within-dyad correlation accounts for reciprocity, i.e., the degree to which relationships are reciprocated. When there is reciprocity, $X_{a,b}$ and $X_{b,a}$ are more likely to be the same. This model is identical to keeping only the multiplicative component of the additive and multiplicative effects (AME) model of \cite{hoff2021additive}. We use $\boldsymbol{U} $ and $\boldsymbol{V} $ to denote the $N \times K$ matrices of the coordinates on the latent network dimensions for sender and receiver, and $\boldsymbol{E}$ to denote the $N \times N$ matrix of network errors. The approximation of the posterior distributions of the unknown quantities is facilitated by setting prior distributions a $N(0, \sigma^{-2}_0)$ with $\sigma^{-2}_0>0$ for $\delta$, a $\text{Unif}(-1, 1)$  for $\rho$, and a $\text{gamma} (1/2, 1/2)$ for $\sigma_e^{-2}$. The prior for the covariance of the latent network dimensions is described in the joint component.

Random node-specific intercepts or node specific additive effects can be added to the above network model \parencite{krivitsky2009representing,hoff2021additive}. \textcite{karrer2011stochastic} makes a case for modeling the heterogeneity of the node degrees with additive node-level (fixed) effects in stochastic blockmodels, resulting in degree-corrected stochastic blockmodels (see discussions about the additive node-level effects in stochastic blockmodels in \textcite{jin2015fast} and \textcite{zhao2012consistency}). Without the additive node-level effects, blockmodels tend to identify clusters based on node degrees (e.g., a cluster of active nodes versus a cluster of less active nodes), especially when there is large heterogeneity in the node degrees. However, it is less clear if the additive node-level effects, e.g. the random row and column intercepts, are needed to model all types of networks in latent variable models. Indeed, the additive effects are useful to capture higher (or lower) nodal activity directed to all nodes in the network irrespective of their positions in the latent space. On the other hand, the length of the latent variable of the node contributes to the activity level of the node in relation to other nodes in the space---this is captured by the multiplicative nature of the vector products to model connections. In small data sets with moderate levels of heterogeneity of the node degrees, the latent positions alone might be sufficient for modeling the activity levels of the nodes. Although, in principle, we can expect improvement in the recovery of the node degrees with added additive effects, a situation might arise where the improvement is negligible. 


In this article, we choose not to include additive effects in the network component of the framework. In our application of the advice-seeking network and the school climate questionnaire, we observe comparable recovery of node degrees as well as comparable model fit with and without the nodal effects in the model. Therefore, we think the nodal effects are not needed in the modeling of the advice-seeking network, though this might not be the case for other networks.

Although the proposed approach for the network component of the joint framework can be seen as an AME model, the difference of our approach is that we connect the network component with the item response component using a covariance-based joint component. See details about the joint component in section \ref{joint}.\\


\noindent \textbf{Identifiability}. The coordinates of the latent network dimensions, $\boldsymbol{U}$ and $\boldsymbol{V}$ are identified up to rotations while $\boldsymbol{U}\boldsymbol{V}^T$ is directly identified following the condition that either $\boldsymbol{U}$ or $\boldsymbol{V}$ is centered. Suppose $\boldsymbol{1}_N$ is the $N$-dimensional column vector of $1$s;  $\boldsymbol{J}_N = \boldsymbol{1}_N \boldsymbol{1}_N^T$ is the $N \times N$ matrix of $1$s; and $\boldsymbol{H}$ is the centering matrix, $\boldsymbol{H} = \boldsymbol{I}_N - \frac{1}{N} \boldsymbol{1}_N \boldsymbol{1}_N^T$. 
\begin{lemma}
Let either $\boldsymbol{U}$ or $\boldsymbol{V}$ be centered, i.e., $
\boldsymbol{H}\boldsymbol{U} = \boldsymbol{U},\text{or     } \boldsymbol{H}\boldsymbol{V} = \boldsymbol{V}$. If two sets of parameters $\delta$, $\boldsymbol{U} $, $\boldsymbol{V} $ and $\delta'$, $\boldsymbol{U'} $, $\boldsymbol{V'} $ lead to the same $E(X_{a,b})$, then $\delta = \delta'$, $\boldsymbol{U} = \boldsymbol{U'} \boldsymbol{O} $ and $\boldsymbol{V} = \boldsymbol{V'} \boldsymbol{O}^{-1} $, where $\boldsymbol{O}$ is a $K \times K$ nonsingular matrix. 
\end{lemma}
 The proof of this lemma and that of Lemma 2 on identifiability of the item response model is in the supplementary materials. The proofs follow similar arguments to those laid out in \cite{zhang2020flexible, huang2022mutually}. To resolve the rotational indeterminacy and reflection of the axes in the latent space, we follow the approach in \textcite{fosdick2015testing, minhas2019inferential,hoff_2015_b, hoff2021additive} and use the first $K$ left and right singular vectors of the posterior means of the vector products as the estimates of $\boldsymbol{U}$ and $\boldsymbol{V}$. In this way, we can obtain estimates of the latent positions following the convergence of the vector products, and not the convergence of the latent positions themselves. The convergence of the vector products will be assessed and reported. The consequence of this approach is that we do not have a posterior distribution of the latent positions, only of the vector products.\\



\noindent \textbf{Binary Network}.\label{link}
The observed advice-seeking network in our application is binary and not continuous.  We model the observed binary edge $X_{a,b}$ as the binary indicator that the latent continuous measure, $\phi_{a,b}$ is bigger than zero, i.e., $X_{a,b} = 1$, when $\phi_{a,b} >0$. For a binary network, $\phi_{a,b}$ replaces $X_{a,b}$ in the proposed methodology in Equation \ref{n1}. The error variance of the latent continuous measure is assumed to be $1$ (as in probit models) to identify the scale of the latent continuous measure.

\subsection{Modeling the Item Response Component}\label{sectionirt}

To model the school climate questionnaire, we propose a latent variable model with vector products for continuous observations. Through this methodology, we aim to get three sets of information: (1) item intercept, (2) item slope, and (3) the latent variable values. The item intercept reflects the average of the responses across all persons for each item. A higher intercept indicates a higher average response. The item slope reflects how well the item differentiates in terms of the latent variable. One unit difference in the latent variable results in a large difference in the probabilities of response when the item slope is large.

Consider the continuous item responses, $\boldsymbol{Y}$,  a $N \times M$ matrix, where $N$ is the number of persons and $M$ is the number of items. The response, $Y_{p,i}$ represents person $p$'s response to item $i$, $p = 1,2,..., N$ and $i = 1,2,..., M$. Let $D$ be the number of dimensions or latent variables in the item responses. The model can be written as:
\begin{align}
Y_{p,i} = \beta_i + \boldsymbol{\alpha}_i^T \boldsymbol{\theta}_p + \epsilon_{p,i},  \qquad   \boldsymbol{\theta}_p \overset{iid}{\sim}  \text{MVN}(0, \Lambda_{\theta}) , \qquad \epsilon_{p,i} \overset{iid}{\sim}  N(0, \sigma^2_{\epsilon}), \label{irt}
\end{align}where $\boldsymbol{\alpha}_i$ and $\boldsymbol{\theta}_p$ are the $D$-dimensional vectors containing the item slopes and the person latent variables of the item responses; $\beta_i$ is the intercept of item $i$; and $\sigma^2_{\epsilon}$ is the error variance for the item responses. As is common in models for item responses, the parameters for the items (including the item slope, $\boldsymbol{\alpha}_i$ and the item intercept, $\beta_i$) are fixed, and the person variable, $\boldsymbol{\theta}_p$ are random. We use $\boldsymbol{\beta} $ to denote the $M \times 1$ vector of item intercepts, $\boldsymbol{A}$ to denote the $M \times D$ matrix of item slopes, $\boldsymbol{\Theta}$ to denote the $N \times D$ matrix of latent variables of item responses, and $\boldsymbol{\Psi}$ to denote the $N \times M$ matrix of item response errors. Let us define $\boldsymbol{\xi}_i$ as a vector of item parameters, $ \boldsymbol{\xi}_i = (\boldsymbol{\alpha}_i^T, \beta_i)^T$. Approximation of the posterior distribution of the item parameters is facilitated by setting a $\text{MVN} (\boldsymbol{\mu}_{\xi}, \boldsymbol{\Sigma}_{\xi}),  \boldsymbol{\mu}_{\xi} = (1,1,...,1,0)^T, \boldsymbol{\Sigma}_{\xi} = \boldsymbol{I}_{D+1} $ prior distribution for $\boldsymbol{\xi}_i$. We set a prior distribution of $\text{gamma } (1/2, 1/2)$ for $\sigma_{\epsilon}^{-2}$. The prior for the covariance of the latent variables of the item responses is described in the joint component. In the case of binary item response, similar to binary network edge value, we can model the binary network response as the binary indicator that the latent continuous measure, $\eta_{p,i}$ is larger than zero, i.e., $Y_{p,i} = 1$, when $\eta_{p,i} >0$.\\


\noindent \textbf{Identifiability}. The latent variables of the item responses and the item slopes, $\boldsymbol{\Theta}$ and $\boldsymbol{A}$ are identified up to rotations while $\boldsymbol{\Theta} \boldsymbol{A}^T$ is directly identified following the condition that either $\boldsymbol{\Theta}$ or $\boldsymbol{A}$ is centered. Following convention, we focus on the condition that $\boldsymbol{\Theta}$ is centered. 

\begin{lemma}
Assume $\boldsymbol{\Theta}$ is centered, i.e., $
\boldsymbol{H}\boldsymbol{\Theta} = \boldsymbol{\Theta}$. If two sets of parameters $\boldsymbol{\beta}$, $\boldsymbol{\Theta} $, $\boldsymbol{A} $ and $\boldsymbol{\beta'}$, $\boldsymbol{\Theta}' $, $\boldsymbol{A}' $ lead to the same $E(Y_{p,i})$, then $\boldsymbol{\beta} = \boldsymbol{\beta'}$, $\boldsymbol{\Theta} = \boldsymbol{\Theta'} \boldsymbol{O} $ and $\boldsymbol{A} = \boldsymbol{A'} \boldsymbol{O}^{-1} $, where $\boldsymbol{O}$ is a $D \times D$ nonsingular matrix.
\end{lemma}
The proof is in the supplementary materials. In addition to the condition identified above, we also fix the latent variables to have unit variances and orthogonality between dimensions following the current factor model (and item response model) literature. Similar to the network component of the JNIRM, we also use the first $D$ left and right singular vectors of the posterior means of the vector products as the estimates of $\boldsymbol{\theta}$ and $\boldsymbol{\alpha}$. In this approach, only the convergence of the vector products is needed for the estimation of the item slopes and the person variables, not the convergence of the item slopes and the latent variables themselves.


\subsection{Modeling the Joint Component}\label{joint}

Recall that the network component of the joint framework is modeled following Equation \ref{n1} and that the item response component of the joint framework is modeled following Equation \ref{irt}. In these two components, we use normal distributions for the network dimensions and the latent variables of the item responses separately. To connect these two sets of variables, we use a multivariate normal distribution with a joint covariance matrix: \begin{align}
     (\boldsymbol{u}_p, \boldsymbol{v}_p, \boldsymbol{\theta}_p)^T & \overset{iid}{\sim} \text{MVN} \left(  \begin{psmallmatrix} \boldsymbol{0} \\
  \boldsymbol{0}\end{psmallmatrix},
  \boldsymbol{\Sigma}_{u\theta} 
  \right),  
  \qquad\boldsymbol{\Sigma}_{u\theta}= \begin{pmatrix} \Lambda_u &\Lambda_{u\theta}^T\\
\Lambda_{u\theta} &\Lambda_{\theta} \end{pmatrix}. \label{model}
\end{align} Approximation of the posterior distribution of $  \boldsymbol{\Sigma}_{u\theta} $ is facilitated by setting a prior distribution of $ \text{Wishart} (\boldsymbol{I}_{2K+D}, 2K+D+2)$. The covariance of the latent network dimensions, $(\boldsymbol{u}_p, \boldsymbol{v}_p)^T$ is the $2K \times 2K$ matrix $\boldsymbol{\Lambda}_{u}$, and the covariance of the latent variables of the item responses is the $D \times D$ matrix $\boldsymbol{\Lambda}_{\theta}$. If the network dimensions are independent, then $\boldsymbol{\Lambda}_{u}$ is a diagonal matrix. If the dimensions of the item responses are independent, then $\boldsymbol{\Lambda}_{\theta}$ is a diagonal matrix. In JNIRM, we can test whether the network and item responses are dependent by testing whether the $D \times 2K$ matrix $\boldsymbol{\Lambda}_{u\theta} = 0.$ See details in the next section. The model is estimated using a Gibbs sampler outlined in section \ref{est}.

\subsection{Advantages}

The advantages of JNIRM can be discussed in three ways: (1) JNIRM as a latent space model for networks and as a latent variable model for item responses, (2) JNIRM as a joint modeling framework, and (3) JNIRM as a covariance-based framework for modeling dependence. As has been mentioned in the introduction, as a covariance-based joint framework, JNIRM is expected to show improvement in parameter estimation over separate network and item response models. We also expect the joint model to outperform the approach that extracts latent factors from one component (either the network or the item responses) in order to use them as regressors for predicting the other in a regression model. This is because unlike that approach, our joint model does not ignore the uncertainty associated with extracting the latent factors. 

JNIRM as a joint model poses a few advantages when compared with ($a$) the network model with observed covariates in which the network is modeled conditional on the observed covariates and ($b$) the social influence model in which attributes are modeled conditional on observed features of the network. In these types of models ($a$ and $b$), the relationship between the network and attributes is one-directional: estimated as the effect of the network on the attributes or the effect of the attributes on the network, when in fact, the relationship is often correlational. In these models, either the networks or the attributes are treated as fixed, though they may not be. There may be interest in predicting the network relations and attributes of a node when both the network information and attribute information are partially missing. In such cases, $a$ and $b$ become obsolete as they assume either the network or the attributes to be fully observed. Furthermore, when either the network or the attributes are not modeled, we do not know the uncertainty of the observations. Lastly, we do not gain strength in estimating one side (either the network or the attributes) when the other side is not modeled.

\subsection{Exploring the Dependence between the Network and Item Responses}

In this section, we discuss how to make inferences regarding the dependence between the network and item responses. Due to the previously described solution for the identification issues, we cannot directly determine uncertainty about the covariance matrix based on the MCMC chain. Instead, following \textcite{fosdick2015testing}, we use the multivariate test of independence and the CCA to study the dependence between the two components of the joint framework. We first discuss the test, and then CCA.

\subsubsection{The Multivariate Test of Independence}\label{test}

The classical multivariate test of independence (see \textcite{anderson1962introduction}) is applied using the estimated network dimensions and the latent variables of the item response. More specifically, the null and the alternative hypotheses of the test are: 
\begin{align}
H_0: \Lambda_{u\theta} = 0 \qquad \text{vs.}   \qquad H_1: \Lambda_{u\theta} \neq 0.
\end{align} 

The test statistic\footnote{The test statistic can also be written as the monotonic increasing function of $v$, i.e, $v^{1/2 N}$} is
$t = \frac{\max_{ \Lambda_u,  \Lambda_{\theta}} L_0(\Lambda_u,  \Lambda_{\theta} | \boldsymbol{u}_p, \boldsymbol{v}_p, \boldsymbol{\theta}_p)
}{
\max_{\boldsymbol{\Sigma}_{u\theta}} L(\boldsymbol{\Sigma}_{u\theta}|\boldsymbol{u}_p, \boldsymbol{v}_p, \boldsymbol{\theta}_p)} =  \left( \frac{
|\hat{\boldsymbol{\Sigma}}_{u\theta}|}{|\hat{\Lambda}_u| \cdot | \hat{\Lambda}_{\theta}|} \right)$, where $L_0$ and $L$ are the likelihood with and without restricting $\Lambda_{u\theta}=0$, and estimates of the covariance matrix are obtained using the estimated latent network dimensions and the latent variables of the item responses. The critical region of the test is $t < t(\alpha)$, where $t(\alpha)$ is the threshold value such that the probability of an observed $t$ is smaller than $t(\alpha)$ with probability $\alpha$.

An important reason for applying this test is that the test is invariant to linear transformations of each set of variables, i.e., the test is invariant under transformations of the form $\boldsymbol{B} \begin{psmallmatrix} \boldsymbol{\boldsymbol{u}_p} \\
  \boldsymbol{\boldsymbol{v}_p}\end{psmallmatrix}$ and $\boldsymbol{B}' \boldsymbol{\theta}_p$, where $\boldsymbol{B} $ and $\boldsymbol{B}'$ are $2K \times 2K$ and $D \times D$ nonsingular matrices.  Recall that the network component and the item response component of JNIRM are invariant under certain transformations of the latent variables. The network component is invariant under transformations of the form: $ \boldsymbol{u}_p^T \boldsymbol{O} (\boldsymbol{v}_p \boldsymbol{O}^{-1})^T$, where $\boldsymbol{O}$ is a $K \times K$ nonsingular matrix, and the item response component is invariant under transformations of the form: $\boldsymbol{\theta}_p^T \boldsymbol{O'} (\boldsymbol{\alpha}_i \boldsymbol{O'}^{-1})^T$, where $\boldsymbol{O'}$ is a $D \times D$ nonsingular matrix. It can thus be shown that transformed $\boldsymbol{u}_p, \boldsymbol{v}_p, \boldsymbol{\theta}_p$ with respect to $\boldsymbol{O}$ and $ \boldsymbol{O'}$ can be seen as transformed $\boldsymbol{u}_p, \boldsymbol{v}_p, \boldsymbol{\theta}_p$ with respect to $\boldsymbol{B}$ and $ \boldsymbol{B'}$---the opposite is not true. Therefore, transformations of $\boldsymbol{u}_p, \boldsymbol{v}_p, \boldsymbol{\theta}_p$ that lead to an invariant posterior of the vector products for JNIRM will also lead to the same test result.

Because estimates of the latent network dimensions and the latent variables of the item responses are used, we expect measurement error due to the estimation process. When applying the multivariate test to assess the independence between attributes and the network, \textcite{fosdick2015testing} found that the error induced by estimating the latent network dimensions using the singular values of the vector products has minimal impact on the power of the test compared to using true values. Using simulations, they showed that little power is lost when the latent network dimensions are estimated, even when the sample size is small.

Our application of the multivariate test is similar to the application in \textcite{fosdick2015testing}, and both use latent variable models to conduct the test of independence between a network and attributes. The differences between their approach and the proposed methodology have been discussed in the introduction.

\subsubsection{Canonical Correlation Analysis}\label{section_cca}

\begin{figure}

\centering
  \includegraphics[scale=0.2]{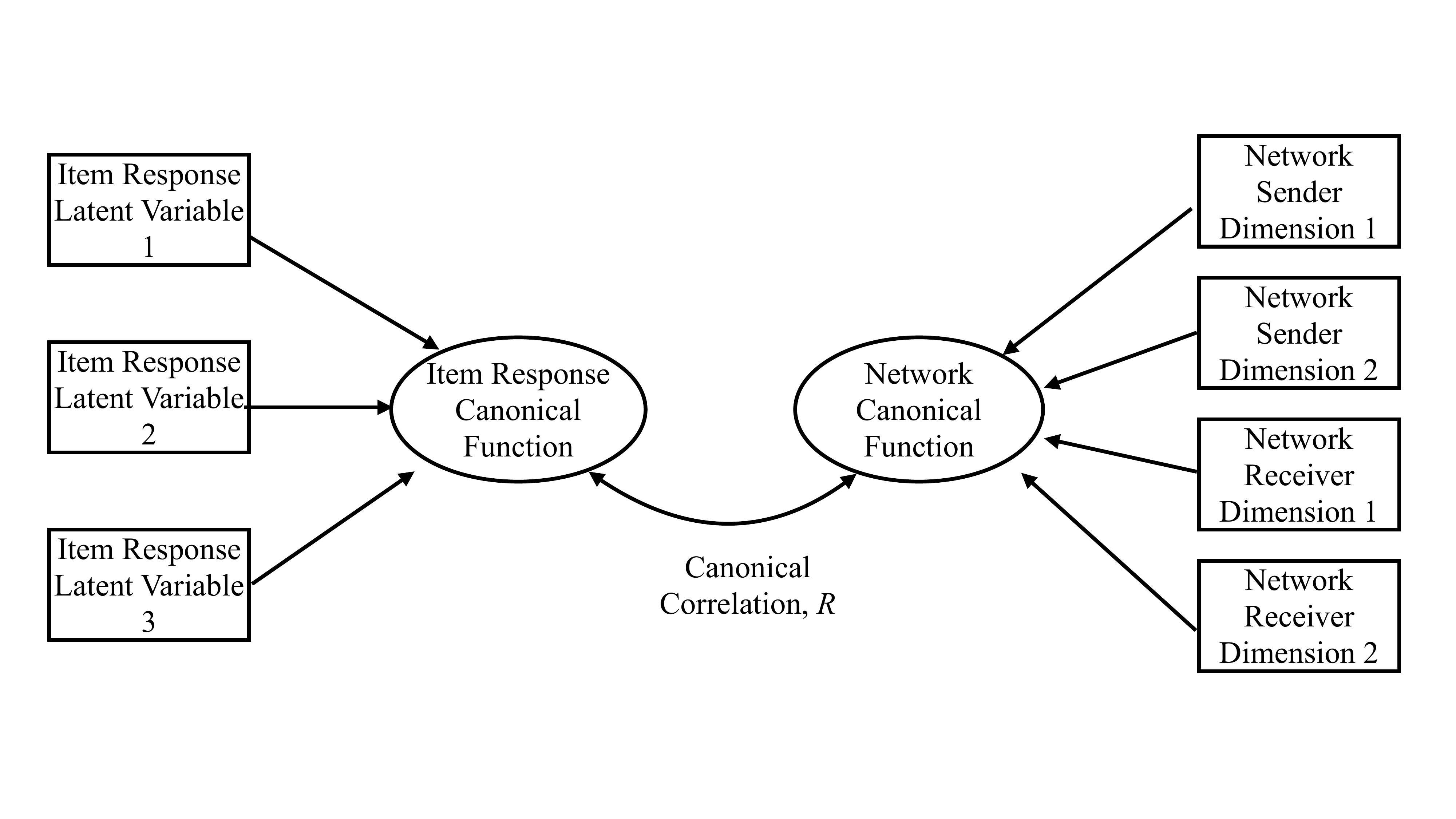}
 \caption{Illustration of the network and item response canonical functions in a canonical correlation analysis with three latent variables from the item responses and two network dimensions. The canonical correlation is the correlation between the network and item response canonical functions, which are linear combinations of the item response latent variables and the network dimensions.}
\label{cancor}
\end{figure}

Following the rejection of the null hypothesis in the classical multivariate test of independence between the network and item responses, we may wish to know where the lack of independence lies. We can use the CCA to achieve this goal. In this section, we provide a brief overview of CCA, and we focus on how it can be applied to understand the relationship between the network and item responses. For more detailed discussions about this topic, we refer readers to \textcite{marden2015multivariate,sherry2005conducting,thompson1984canonical,anderson1962introduction}.

CCA is a useful technique to study the relationship between two sets of variables. In this case, the first set of variables includes the network dimensions from JNIRM, and the second set of variables includes the latent variables from the item response component of JNIRM. Using CCA, we estimate the largest correlation obtainable between a linear combination of the network dimensions and a linear combination of the latent variables of the item responses. Then, a second largest correlation is obtained with linear combinations that are uncorrelated with the first linear combinations. The process is repeated, with each set of linear combinations maximizing the correlation subject to being uncorrelated with the previous combinations.

In Figure \ref{cancor}, we demonstrate CCA with a hypothetical JNIRM when the number of dimensions for the network is two, and the number of dimensions for the item responses is three. In the left side of the figure, the latent variables of the item responses are linearly combined to create the item response canonical function, and a similar process occurs in the right side of the figure for the network. The weights of the linear combination are called the canonical weights or the function coefficients, and the resulting weighted sum is called the canonical function. Function coefficients can be standardized and unstandardized just as regression coefficients. For example, an item response canonical function is the weighted sum of the latent variables of the item responses.  There can be as many canonical functions as the smallest set of variables at one of the two sides.  The canonical functions are orthogonal to each other following the independence of the linear combinations. We will base our interpretations with CCA on standardized function coefficients.


In the center of Figure \ref{cancor}, the canonical correlation, $R$ is shown as the correlation between the network and the item response canonical functions. The square of the canonical correlation, $R^2$ represents the proportion of variance shared by the network and the item response canonical functions. The square of the second canonical correlation indicates the proportion of the shared variance between the second pair of canonical functions based on the remaining relationship (independent of the first pair of canonical functions).  


Significance tests can be conducted in association with the canonical correlations. Testing each correlation separately for statistical significance is not readily available, instead, the correlations are tested in a hierarchal fashion. First, we test whether all correlations are zero; if that null hypothesis is rejected, we can test whether the $I-1$ ($I$ is the total number of canonical correlations) smallest correlations are---or whether the remaining shared variance is---zero, etc.

\section{Bayesian Inference}\label{est}

In this section, we propose an algorithm for estimating the proposed joint network and item response model (JNIRM) using Markov chain Monte Carlo (MCMC) methods. 
We devise a
Gibbs sampler for the proposed model, and we briefly discuss how to model the binary data. In the following, we provide details of the full conditionals for each unknown quantity, and the algorithm is implemented through iterative sampling of the unknown quantities from their full conditional distributions. With sufficient iterations, we obtain stable Markov chains to approximate various quantities of the targeted posterior distributions. With random initial values on the unknown parameters, we conduct posterior computation by iterating the following steps.

\begin{itemize}
    \item simulate $\boldsymbol{U}, \boldsymbol{V}$ from their full conditional distributions. 
        \item simulate $\boldsymbol{\Theta}$ from their full conditional distributions. 
    \item simulate $\boldsymbol{\Sigma}_{u \theta}$ from its full conditional distribution. 
\item simulate $\sigma^{-2}_e$, $\rho$ and $\delta$ from their full conditional distributions. 
\item simulate $\sigma^{-2}_{\epsilon}$ from its full conditional distributions. 
\item simulate $\boldsymbol{\xi}_i$ from its full conditional distributions. 

\end{itemize}

\subsection{Full Conditionals of the Latent Variables and the Covariance}

 \subsubsection{Multiplicative UV Effect}

The item responses are related to the network through the dependence between $\boldsymbol{\Theta}$ and $\boldsymbol{U}, \boldsymbol{V}$. We are interested in the joint full conditional distribution of $\boldsymbol{\Theta}$ and $\boldsymbol{U}, \boldsymbol{V}$. For person $p$, we focus on the first dimension for sender, $u_{p,1}$ conditional on the other dimensions, $u_{p,2},...,u_{p,K}$ and all dimensions of $\boldsymbol{v}_p $, $\boldsymbol{v}_p = (v_{p,1},...,v_{p,K})^T$. Given that $\boldsymbol{u}_p, \boldsymbol{v}_p$
and $\boldsymbol{\theta}_p$ follow a multivariate normal distribution, the joint probability model for $u_{p,1}$ and $\theta_{p,1}$, ..., $\theta_{p,D}$ conditional on $u_{p,2},...,u_{p,K}$ and $\boldsymbol{v}_p$ is also a multivariate normal distribution with conditional mean $\boldsymbol{\mu}_{u\theta|...}$ and conditional covariance matrix $\boldsymbol{\Sigma}_{u\theta|...}$. We can define the conditional covariance matrix as a block matrx: $
\boldsymbol{\Sigma}_{u\theta|...} = \begin{pmatrix} \Lambda_u &\Lambda_{u\theta}^T\\
\Lambda_{u\theta} &\Lambda_{\theta} \end{pmatrix}, \nonumber
$ and the inverse of this block matrix is \begin{align}
\boldsymbol{\Sigma}_{u\theta|...}^{-1} = 
\begin{pmatrix} \Lambda_u &\Lambda_{u\theta}^T\\
\Lambda_{u\theta} &\Lambda_{\theta} \end{pmatrix}^{-1} 
= \begin{pmatrix} (\Lambda_u - \Lambda_{u\theta}^T \Lambda_{\theta}^{-1} \Lambda_{u\theta})^{-1} &
- (\Lambda_u - \Lambda_{u\theta}^T \Lambda_{\theta}^{-1} \Lambda_{u\theta})^{-1} \Lambda_{u\theta}^T \Lambda_{\theta}^{-1}\\
- (\Lambda_{\theta} - \Lambda_{u\theta} \Lambda_u^{-1} \Lambda_{u\theta}^T)^{-1} \Lambda_{u\theta} \Lambda_u^{-1} &(\Lambda_{\theta} - \Lambda_{u\theta} \Lambda_u^{-1} \Lambda_{u\theta}^T)^{-1} \end{pmatrix}. \label{lam2}
\end{align} We can further define the inverse of the conditional covariance matrix as another block matrix:
$
\boldsymbol{\Sigma}_{u\theta|...}^{-1} = \begin{pmatrix} Q_u &Q_{\theta u}\\ 
Q_{u\theta} & Q_{\theta}\end{pmatrix},
$ with each component as a function of $\Lambda$s.  Therefore, the probability model for $u_{p,i}, \theta_{p,1},...,\theta_{p,D}$ conditional on $u_{p,2},.., u_{p,K}, v_{p,1},...,v_{p,K}$  can be written as 
 \begin{align}
p\left(\begin{psmallmatrix} u_{p,1} \\\boldsymbol{\theta}_p\end{psmallmatrix} | ...\right) \propto  &\exp \left(  -\frac{1}{2} \left( \begin{psmallmatrix}u_{p,1} \\ \boldsymbol{\theta}_p\end{psmallmatrix}- \boldsymbol{\mu}_{u\theta |...} \right)^T
 \begin{psmallmatrix} Q_u &Q_{\theta u}\\ 
Q_{u\theta} & Q_{\theta}\end{psmallmatrix}
 \left( \begin{psmallmatrix} u_{p,1} \\ \boldsymbol{\theta}_p \end{psmallmatrix} - \boldsymbol{\mu}_{u\theta |...} \right) \right)\nonumber\\
  \propto &
\exp \left(-\frac{1}{2}  (u_{p,1}^T Q_u u_{p,1} + u_1^T Q_{\theta u} \boldsymbol{\theta}_p +u_{p,1}^TQ_{u \theta}^T\boldsymbol{\theta}_p
  )
  +  u_{p,1}^T S_u  \right), 
\end{align} where $S_u$ is the first block of the block matrix $ \begin{psmallmatrix} S_u\\ 
S_{\theta}\end{psmallmatrix}_{(D+1) \times 1} = \begin{psmallmatrix} Q_u &Q_{\theta u}\\ 
Q_{u\theta} & Q_{\theta}\end{psmallmatrix}  \boldsymbol{\mu}_{u\theta |...}. $ 

Let us now look at the first dimension for all persons and rewrite the network component as:  $\boldsymbol{R} = \boldsymbol{X}- \delta \boldsymbol{1} \boldsymbol{1}^T - \sum_{k=2}^K \boldsymbol{u}_k \boldsymbol{v}_k^T $, where $\boldsymbol{u}_k$ is the $N \times 1$ vector representing the $k$th network dimension for sender, and $\boldsymbol{v}_k$ is the $N \times 1$ vector representing the $k$th network dimension for receiver. Then, we have $\boldsymbol{R} = \boldsymbol{u}_1^T \boldsymbol{v}_1 + \boldsymbol{E}$. Decorrelating the error, we have $\boldsymbol{\tilde{R}} = c \boldsymbol{R} + d \boldsymbol{R}^T$, where $
c= \sigma_e^{-1} ((1+\rho)^{-1/2} + (1- \rho)^{-1/2})/2$, and $
 d=  \sigma_e^{-1} ((1+\rho)^{-1/2} - (1- \rho)^{-1/2})/2$. We can write the model for $\boldsymbol{\tilde{R}}$ in a vectorized form and obtain the likelihood for the decorrelated network as $
f(\boldsymbol{\tilde{r}} | \boldsymbol{v}_1, \boldsymbol{u}_1) \propto  \exp \left( - \frac{1}{2} 
\left( \boldsymbol{\tilde{r}} - \boldsymbol{M} \boldsymbol{u}_i \right)^T\left( \boldsymbol{\tilde{r}} - \boldsymbol{M} \boldsymbol{u}_i \right)
\right)$, $\boldsymbol{M} = c(\boldsymbol{v}_1 \otimes \boldsymbol{I}) + d(\boldsymbol{I} \otimes \boldsymbol{v}_1)$. 

We can write the full conditional distribution of $\boldsymbol{u}_{1}$:  \begin{align}
&f(\boldsymbol{u}_1| \boldsymbol{\theta}_1,\boldsymbol{\theta}_2,...,\boldsymbol{\theta}_D,  \boldsymbol{u}_2,..,\boldsymbol{u}_K, \boldsymbol{v}_1,..,\boldsymbol{v}_K) \nonumber \\
&\propto
f(\boldsymbol{\tilde{r}} | \boldsymbol{v}_1, \boldsymbol{u}_1) \prod_{p=1}^N p\left(\begin{psmallmatrix} u_{p,1} \\\boldsymbol{\theta}_{p}\end{psmallmatrix} | ...\right) \nonumber \\
  & \propto  \exp \left( - \frac{1}{2}  \boldsymbol{u}_1^T \boldsymbol{M}^T\boldsymbol{M} \boldsymbol{u}_1 +  \boldsymbol{u}_1^T \boldsymbol{M}^T \boldsymbol{\tilde{r}}
  -\frac{1}{2}  \sum_{p=1}^N (u_{p,1}^T Q_u u_{p,1} + u_{p,1}^T Q_{\theta u} \boldsymbol{\theta}_p + u_{p,1}^TQ_{u \theta}^T\boldsymbol{\theta}_p
  )
  +   \sum_{p=1}^N u_{p,1}^T S_u  \right).
\end{align} The full conditional distribution of the other network dimensions and the latent variables of item responses can be derived in a similar fashion.

Recall that prior for $\boldsymbol{\Sigma}_{u \theta}$ is $\boldsymbol{\Sigma}_{u \theta}^{-1} \sim \text{Wishart} (\boldsymbol{I}_{2K+D}, 2K+D+2)$. Let $\boldsymbol{F'}$ be a $N \times (2K+D)$ matrix with $p$th row as $(\boldsymbol{u}_{p}^T, \boldsymbol{v}_{p}^T, \boldsymbol{\theta}_{p}^T)$. The full conditional for $\boldsymbol{\Sigma}_{u \theta}$ follows a inverse-Wishart $(\boldsymbol{I}_{2K+D} + \boldsymbol{F'}^T \boldsymbol{F'}, N+2K+D+2)$.

\subsection{Full conditionals of $\sigma^{-2}_e$, $\rho$ and $\delta$. }
To derive the full conditional distribution of the network error term, we look at the network model with only the error term: 
$\boldsymbol{E} = \boldsymbol{X} - \delta \boldsymbol{1} \boldsymbol{1}^T  - \boldsymbol{U}^T \boldsymbol{U}$ for the bilinear effect, and $\boldsymbol{E} = \boldsymbol{X} - \delta \boldsymbol{1} \boldsymbol{1}^T  - \boldsymbol{U}^T \boldsymbol{V}$ for the multiplicative UV effect. Recall that the prior for $\sigma_e^{-2}$ is gamma $(1/2, 1/2)$. Therefore, the full conditional of $\sigma_e^{-2}$ is also gamma \\$\left(\frac{N^2 + 1}{2}, 1/2 \left( 1+
\sum_{a < b}  \begin{pmatrix} e_{a,b}\\ e_{b,a}  \end{pmatrix}^T  \begin{pmatrix} 1 & \rho \\ \rho & 1   \end{pmatrix}^{-1}  \begin{pmatrix} e_{a,b}\\ e_{b,a}  \end{pmatrix}
+ \sum_{a =1}^N (1 + \rho)^{-1} e_{a,a} ^2
\right)\right)$.

For $\rho$, the full conditional distribution of the within-dyad correlation is not a well-known distribution.  To update parameter $\rho$, we use Metropolis–Hastings with a uniform prior between $-1$ and $1$. We approximate the full conditional distribution by sampling values from a proposed probability model and accept with some probability. The estimation of the intercept parameter, $\delta$, can be seen as a Bayesian linear regression problem with a design matrix $\boldsymbol{1} \boldsymbol{1}^T $ and a normal prior.

\subsection{Full conditionals of item parameters and $\sigma^{-2}_{\epsilon}$.}
Recall that $\boldsymbol{\xi}_i$ is the $(D +1) \times 1$ vector of parameters for item $i$, $\boldsymbol{\xi}_i = (\boldsymbol{\alpha}_i^T, \beta_i)^T$. To derive the full conditional distributions of the item parameters, we rewrite the model for the $i$th item as:  \begin{align}
\boldsymbol{y}_i = \boldsymbol{G} \boldsymbol{\xi}_i + \boldsymbol{\epsilon}_i, 
\end{align}where $\boldsymbol{y}_i$ is the $i$th column vector of the item response matrix; $\boldsymbol{G}$ is a $N \times (D + 1)$ design matrix,  $\boldsymbol{G} = (\boldsymbol{\Theta},  \boldsymbol{1}_{N \times 1})$; and $\boldsymbol{\Theta}$ is the $N \times D$ matrix of the latent variables of the item responses. The error of the item response model follows a normal distribution, $\epsilon_{p,i} \overset{iid}{\sim}  N(0, \sigma^2_{\epsilon})$.

Recall that the prior for $\boldsymbol{\xi}_i$ is a multivariate normal distribution, $\boldsymbol{\xi}_i \sim N (\boldsymbol{\mu}_{\xi}, \boldsymbol{\Sigma}_{\xi})$,  then the full conditional of $\boldsymbol{\xi}_i$ also follows a multivariate normal distribution with mean $\boldsymbol{\mu}_{\xi}^*$ and variance $\boldsymbol{\Sigma}_{\xi}^*$: 
$\boldsymbol{\mu}_{\xi}^* = \Sigma_{\xi}^* \left(  \sigma^{-2}_{\epsilon} \boldsymbol{G}^T \boldsymbol{\eta}_i + \boldsymbol{\Sigma}_{\xi}^{-1} \boldsymbol{\mu}_{\xi}  \right)$, 
$ \Sigma_{\xi}^* =\left( \sigma^{-2}_{\epsilon} ( \boldsymbol{G}^T  \boldsymbol{G}) + \boldsymbol{\Sigma}_{\xi}^{-1}
 \right)^{-1}$.

 The prior for $\sigma^{-2}_{\epsilon}$ is gamma (1/2, 1/2). Therefore, the posterior for  $\sigma^{-2}_{\epsilon}$ is also gamma $\left( \frac{N*M +1}{2}, 1/2*(1 + \sum_{p=1}^N \sum_{i=1}^M \epsilon_{p,i}^2 ) \right)$.

For binary data, $X_{a,b}$ and $y_{p,i}$ are functions of the latent responses, $\phi_{a,b}$ and $\eta_{p,i}$. The data can be seen as binary indicators of the latent responses such that $X_{a,b} = 1$ if $\phi_{a,b} > 0$ and  $Y_{p,i} = 1$ if $\eta_{p,i} > 0$.  The error variances of the latent responses, $\sigma^2_e$ and $\sigma^2_{\epsilon}$ are assumed to be $1$.

\section{School Climate and Teachers' Advice-seeking Network}

We apply the model to each of the 14 schools separately. There are statistically significant relationships between the network and item responses in three out of the fourteen schools. The power of the test is small because of the small sample size per school. We extensively investigate the results for one of the three schools, and we discuss the results of the other two schools to evaluate the generalizability of the relationship between the network dimensions and the questionnaire dimensions. We also provide a summary of the relationships across all schools.

\textbf{Estimation.}
Because the structures of the advice-seeking networks are distinct across different schools, we fitted the JNIRM to each school separately.  To be succinct in our report, we chose one school, school 56, as a more detailed demonstration and summarized the results of the other schools. See discussions about the different network structures recovered for different schools and detailed reports for applications with multiple schools and other individual schools in the supplementary material. There were a total of $26$ teachers in school 56, and the density of the directed network is $0.143$. The receiver and the sender degrees had a negative correlation of $-0.514$, suggesting that, in this school, teachers who often gave advice tended not to seek advice, and vice versa. Note that this finding is not replicated in every school, an example of the difference in network structures in different schools. 
The MCMC algorithm described in section \ref{est} was used to obtain the posterior distributions. We generated $200,000$ iterations and discarded the first $10,000$ as burn-in. We used the trace plots as well as the Gelman and Rubin's diagnostics as the convergence criteria, and neither showed obvious signs of non-convergence (see details in the supplementary material).

\begin{figure}

\centering
  \includegraphics[scale=0.2]{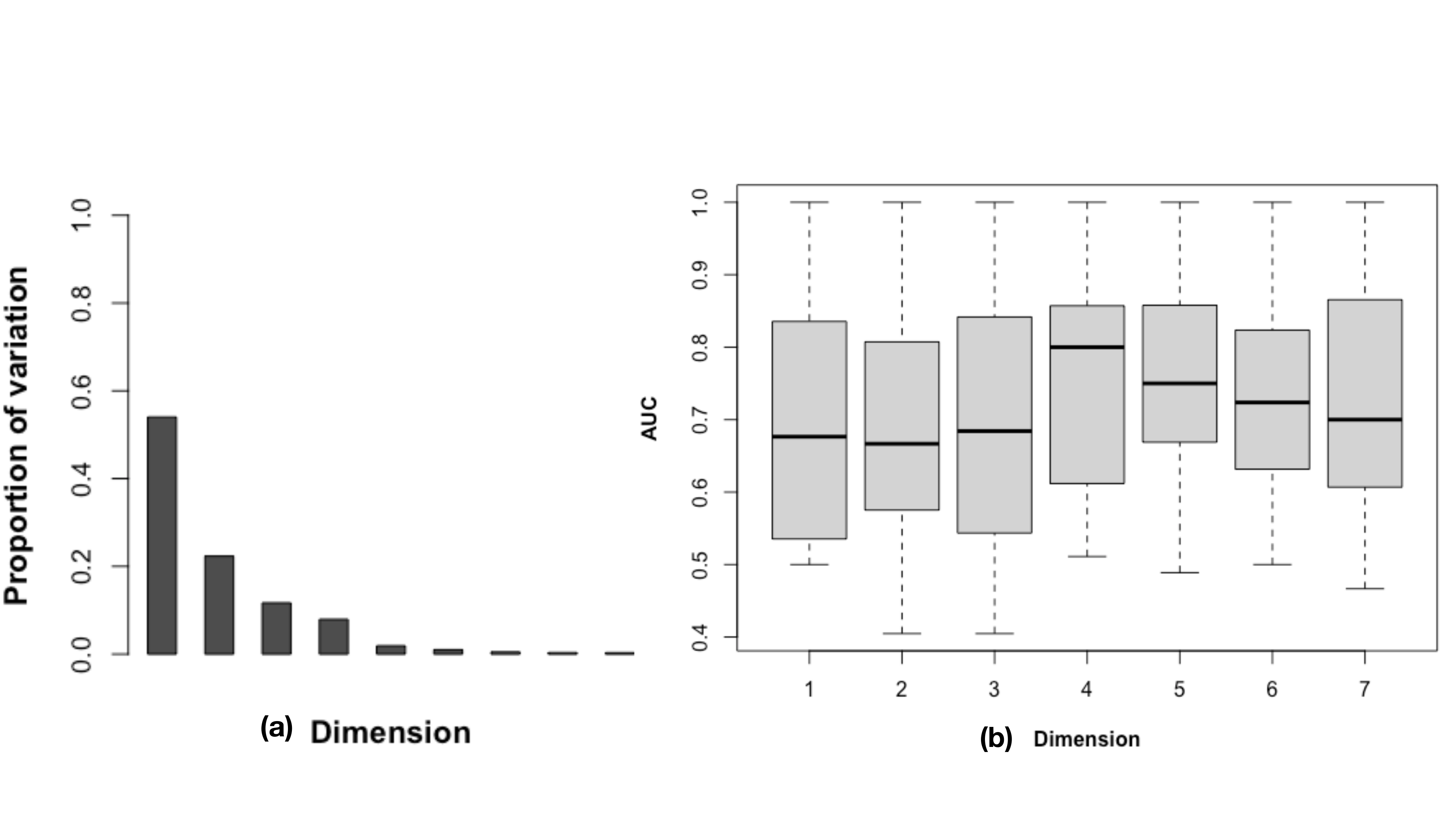}
 \caption{ (a) The proportion of variation explained by each dimension for school 56. (b) The out-of-sample AUC values on the test data when the number of dimensions are from 1 to 7.}
\label{dim}
\end{figure}

\textbf{Dimensionality of the Network.}\label{dimension}
We present two methods for choosing the number of latent dimensions in the network. For the first method, we compared the relative proportion of variation explained by each dimension following \textcite{fosdick2015testing}. This approach is similar to the scree plot method commonly seen in PCA  except that each dimension is estimated based on the proposed joint framework instead of the PCA. Recall that the number of dimensions in the item responses is chosen as 3 based on the number of theoretical constructs, and therefore does not need a data-driven solution. Figure \ref{dim} (a) presents the proportion of variation in $\boldsymbol{\hat{U}} \boldsymbol{\hat{V}}^T$ explained by each dimension of the JNIRM with $K=9$ and $D=3$ fitted to the advice-seeking network and the school climate questionnaire. The results show that there are likely $4$ dimensions for the network component of the model given the big drop in the proportion of variation explained after 4 latent components.  For the second method, we evaluated the out-of-sample model fit under different numbers of dimensions, $K \in {2,3,4,5,6,7}$. We randomly selected one row of the social network (seeking advice) as the test data and the other rows as the training data to obtain the out-of-sample predictive fit, and we repeated the process for each row of the network. The model uses the node's receiving behaviors and the item responses to perform the out-of-sample prediction for each node's seeking behaviors. Figure \ref{dim} (b) shows the out-of-sample AUC values on the test data when the number of dimensions ranges from 1 to 7. The results show that the fit of the JNIRM with $K=4$ was the most optimal with the highest AUC values. Based on the convergent results of both methods, we believe that four is the best choice for the number of network dimensions.

\textbf{Model Fit.}
The joint model was estimated with the network component defined in Equation \ref{n1} using $K=4$, the item response component defined in Equation \ref{irt} using $D=3$ and the joint component defined in Equation \ref{model}. As mentioned before, the random intercepts were not included in the model given that the recoveries of the node degrees were comparable with and without random intercepts in the model (see supplementary materials), and no substantial improvement in model fit was observed when random intercepts were included. The network dimensions for sender and receiver, $\boldsymbol{U}$ and $\boldsymbol{V}$, were estimated by the first $4$ left and right singular vectors of the posterior mean of the vector product, $\boldsymbol{U} \boldsymbol{V}^T$. The latent variables for the item responses and the item slopes, $\boldsymbol{\Theta}$ and $\boldsymbol{A}$, were estimated by the first $3$ left and right singular vectors of the posterior mean of the vector product, $\boldsymbol{\Theta} \boldsymbol{A}^T$.  We performed a target rotation on the estimated item slopes using the same procedure as described before. The rotated slopes again show clear correspondence with the theoretical structure of the dimensions of satisfaction, the perceived influence over school policies, and the perception of students. See details of the rotation and information about other parameter estimates in the supplementary material.  

For the advice-seeking network, the in-sample AUC and the median out-of-sample AUC were estimated as $0.998$ and $0.800$, showing a reasonable fit. In addition, we assessed the fit of the joint model to the data using posterior predictive checking. More specifically, we assessed the posterior predictive fit using $200,000$ replicated data sets based on the estimated model parameters from the MCMC chain. If the estimated statistics based on the replicated data recover the observed statistics in the observed data, it shows that the model fits the data reasonably well. To investigate the fit of the joint model, we assess the recovery of the node degrees, mean item responses per item and per response category and the item covariances, which all show satisfactory fit of the model (details are in the supplementary material).

\begin{figure}

\centering
  \includegraphics[scale=0.2]{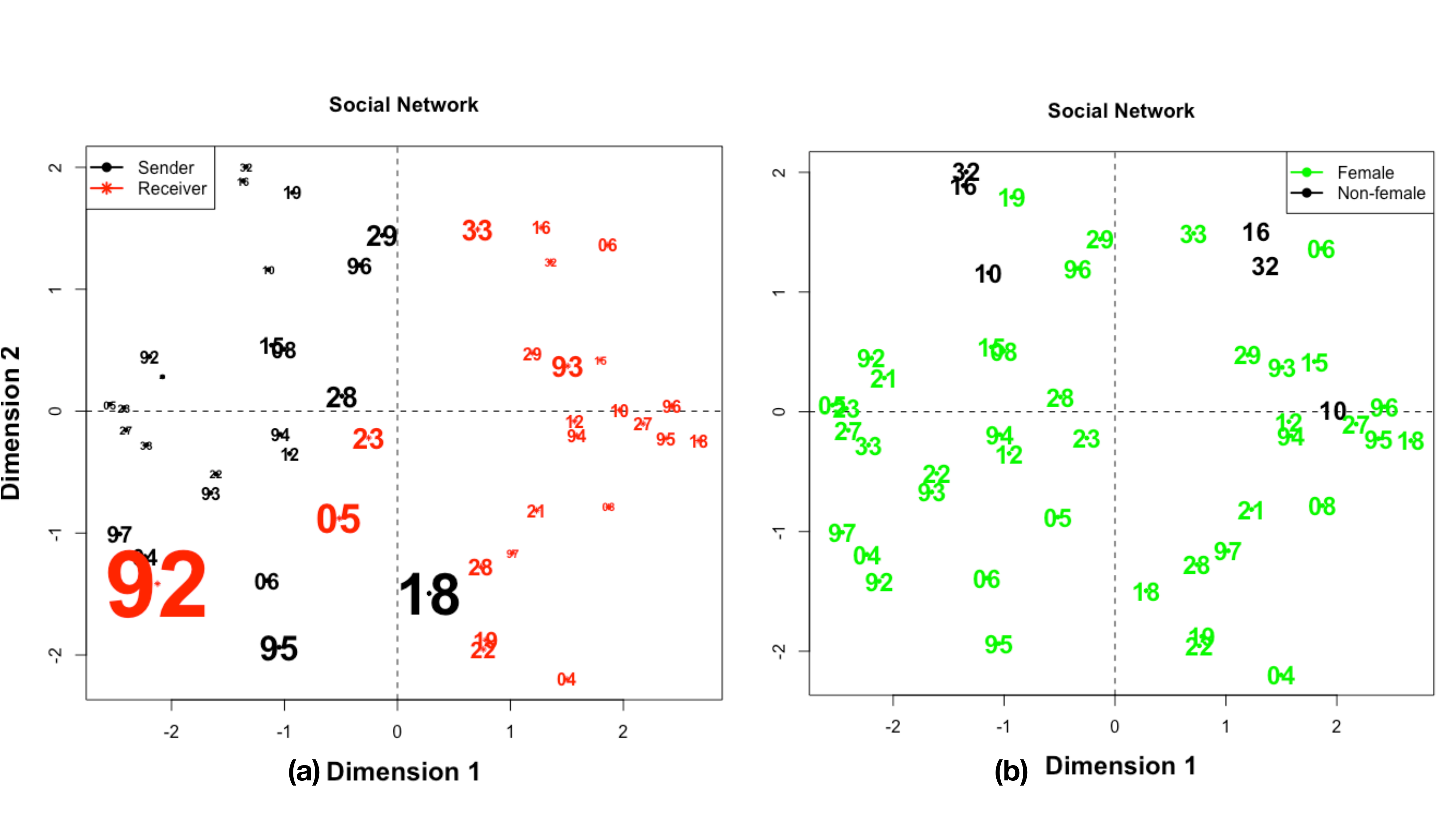}
 \caption{ (a) The estimated latent positions for the first and second dimensions of the latent space. The teachers' IDs are shown as numbers in the plots. The physical size of the ID reflects the node degree. The colors differentiate between sender (black) and receiver (red). (b) The same plot, except that the colors differentiate whether the nodes identify as women, and size of the numbers is constant.}
\label{latentnetwork}
\end{figure}

 \setlength{\arrayrulewidth}{.3mm}
\setlength{\tabcolsep}{10pt}
\renewcommand{\arraystretch}{.4}
\begin{table}[!ht]\caption{Correlations between Network Latent Dimensions and Node Degrees}
 \begin{center}
\begin{tabular}{lccc} 
\toprule
&& \multicolumn{2}{c}{Node Degree} \\
 \cmidrule(lr){3-4}
Latent Dimension & Dimension & Receiver & Sender\\
 \midrule
\multirow{4}{*}{\bfseries Sender} & $D=1$ & -0.413     &    0.685\\
& $D=2$ &0.079    &    -0.424    \\
 &$D=3$& 0.125     &   -0.521   \\
  &$D=4$& 0.071    &    -0.035  \\
  \addlinespace
  \multirow{4}{*}{\bfseries Receiver} & $D=1$ & -0.923      &   0.584\\
& $D=2$ &-0.249   &      0.009   \\
 &$D=3$& -0.375       &  0.016   \\
  &$D=4$& 0.143      &   0.087 \\
\bottomrule
\end{tabular}
\label{correlations}
\end{center}
\end{table} 
\textbf{Interpretation of the Network Latent Space.}
 We first looked at the relationships between the estimated network dimensions and the observed node degrees. Table \ref{correlations} presents the correlations between the estimated network dimensions and the observed sender and receiver degrees. Figure \ref{latentnetwork}a displays the estimated latent positions in the latent subspace based on the first two dimensions. The teachers' IDs are shown as the numbers in the plots. The physical size of the ID reflects the node degrees. For example, teacher 92 was the most active advice giver, and teacher 18 was the most active advice seeker. The colors differentiate between sender (black) and receiver (red). In Table \ref{correlations} and Figure \ref{latentnetwork}, we can see that the sender and receiver coordinates on the first dimension were highly correlated with the node degrees: $0.685$ (sender degree) and $-0.923$ (receiver degree).  The first dimension also differentiated senders from receivers with receivers at the positive side of the x axis and senders at the negative side of x axis. In this way, the first dimensions not only captured the sender and receiver degrees, but also the negative correlations between the sender and receiver node degrees. In this school, teachers who often gave advice tended not to seek advice, and this phenomenon was captured by the first dimensions. In Table \ref{correlations}, the sender coordinates on the second dimension showed moderate negative correlations with the sender degrees, unlike those on the first dimension, which seemed to be unrelated to the receiver degrees.

In addition to the relationships with node degrees, the latent space can also reflect network homophily, which means that the relationships of teachers with each other depend on attributes such as gender, educational background, personality, cultural heritage, etc. In Figure \ref{latentnetwork}b, we present the first and the second dimensions of the latent space colored by whether the teachers identified as female. It seems that those who did not identify as female were positioned higher on the second dimension of the latent space than the female teachers. Therefore, gender could potentially serve as an example from a myriad of possibilities why teachers sought advice from some teachers but not from others.

 \setlength{\arrayrulewidth}{.3mm}
\setlength{\tabcolsep}{10pt}
\renewcommand{\arraystretch}{.4}
\begin{table}[!ht]\caption{Canonical Correlation Results for the First Canonical Functions}
 \begin{center}
\begin{tabular}{lccc} 
\toprule
 Latent Variable & Standardized function coefficients \\
 \midrule
Network Sender 1 & \underline{-0.912}  \\
Network Sender 2 & \underline{0.984} \\
Network Sender 3 & 0.021  \\
Network Sender 4 & 0.040 \\
  \addlinespace
Network Receiver 1 & \underline{1.007} \\
Network Receiver 2 & -0.277 \\
Network Receiver 3 & -0.326\\
Network Receiver 4 & 0.262 \\
  \addlinespace
  Student& 0.150 \\
    Policy& 0.100 \\
      Satisfaction& \underline{-0.889} \\
\bottomrule
\end{tabular}
\label{cca}
\end{center}
\end{table}


\textbf{Relationship between Network and Item Responses.}\label{56re}
The test of independence rejects of the null hypothesis of independence between the network and item responses with Wilks' $\lambda$ of $0.084$ and p-values of $0.009$. Subsequently, we performed the sequential hypothesis tests for the canonical correlation coefficients. There are a total of three pairs of canonical functions. We performed two sequential tests assessing (1) the null hypothesis that the remaining relation after removing the first pair of canonical functions is zero, and (2) the null hypothesis that the remaining relation after removing the first and second pair of canonical functions is zero. The first test resulted in a Wilks’ $\lambda$ of $0.299$ and a p-value of $0.124$, and the second test resulted in a Wilks’s $\lambda$ of $0.656$ and a p-value of $0.312$. Therefore, we failed to reject both null hypotheses, and we focused on interpreting the first canonical correlation. 

Table \ref{cca} presents the CCA results associated with the first canonical correlation. In the table, standardized canonical function coefficients are presented. For emphasis, coefficients for selected latent variables with large coefficients (in absolute values) were underlined. Regarding the item responses, the third dimension of the item responses was the primary contributor to the first canonical correlation. The third dimension of the item responses had a large negative weight in the first canonical function. Recall that the third dimension was the satisfaction dimension. For the network, there were three large weights for the first canonical function: Sender 1 had a large negative weight; Sender 2 and Receiver 1 had large positive weights. Therefore, the satisfaction dimension was positively related to Sender 1, and negatively related to Sender 2 and Receiver 1. Sender 1 was highly positively correlated with sender degree ($0.685$), and it had a high negative weight in the canonical function ($-0.912$). Satisfaction had a high negative weight as well ($-0.889$). We can thus conclude that there was a positive relationship between being satisfied with the school and the frequency of seeking advice. Receiver 1 was highly negatively correlated with receiver degree ($-0.923$), and it had a high positive weight in the canonical function ($1.007$). We can also conclude that there was a positive relationship between being asked for advice and being satisfied with the school. Sender 2 showed similar relationships. Sender 2 was negatively correlated with sender degree ($-0.424$), and it had a high positive weight in the canonical function.  Both asking for advice and being asked for advice seemed to be positively associated with being satisfied with the school.

In Table \ref{cca}, Sender 1 and Receiver 1 had function coefficients with opposite signs---senders high on the first dimension of the network were more likely to have positive reports of satisfaction (high on the satisfaction dimension); and receivers high on the first dimension of the network were more likely to have less positive reports of satisfaction (low on the satisfaction dimension). Thus in school 56, the complementarity principle applies, and teachers get in touch with those who score in an opposite way on the satisfaction dimension. The complementarity principle can also be observed in school 24, where teachers get in touch with those who score in an opposite way on the student dimension (see the supplementary materials)

\begin{figure}

\centering
  \includegraphics[scale=0.12]{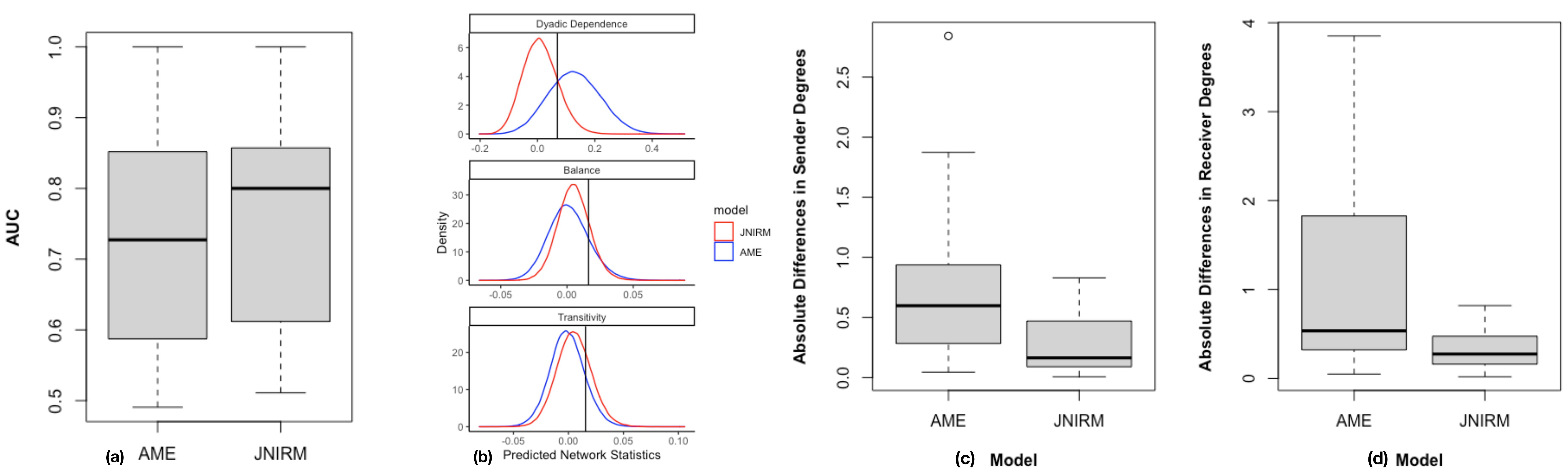}
 \caption{ (a) The out-of-sample AUC values using the separate AME model versus the JNIRM.  (b) The density of three network statistics based on the replicated data using AME versus JNIRM. The vertical lines represent the network statistics in the observed data. The box plots of the absolute differences between the predicted node degrees and the observed node degrees based on AME versus JNIRM. (c)The box plots of the absolute differences between the predicted sender degrees and the observed sender degrees based on AME versus JNIRM. (d)The box plots of the absolute differences between the predicted sender degrees and the observed sender degrees based on AME versus JNIRM.}
\label{aucame}
\end{figure}

\textbf{Comparison to Separate Network and Item Response Models.}\label{56com} 
The rejection of the null hypothesis of independence between the network and item responses for school 56 suggested that the item responses were relevant information for the network, and vice versa.  With the incorporation of relevant new information, we expect an improvement in model fit to data using a joint modeling framework over modeling the data sets separately given that reasonable models are specified on both sides and that there is dependence. In this section, we assess whether there is improvement in model fit using the joint framework compared with the separate models for network and item responses using accuracy of out of sample prediction. We fitted the advice-seeking network to the same network model in Equation \ref{model} with $K=4$ and no item responses using the amen package \parencite{hoff2015dyadic}. Figure \ref{aucame}a shows the out-of-sample AUC values using the separate AME model versus JNIRM. The median out-of-sample AUC value was $0.800$ with JNIRM versus $0.727$ with AME suggesting an improvement in model fit with the joint modeling framework. In Figure \ref{aucame}, we also compare the recovery of network statistics such as node degrees, dyadic dependence, transitivity and balance based on the replicated data sets. The replicated data sets were obtained for posterior predictive checking in the same way as before. For the recovery of the node degrees, the differences were smaller based on both the sender and the receiver degrees using JNIRM versus AME (see Figures \ref{aucame}c and \ref{aucame}d). For the recovery of dyadic dependence and balance, the distributions based on JNIRM were narrower than the distributions based on AME suggesting that there was more precision using JNIRM than AME.

To compare the joint framework with the separate model for the item responses, we also fitted the school climate item responses to the same item response model in Equation \ref{model} with $D = 3$ and no advice-seeking network. We compared the average predicted item responses across replicated data per observed category and per item in addition to the recovery of the item covariances. We found that there was no noticeable difference between the joint and the separate model in the average predicted item responses (see supplementary material for details). 
On the other hand, the joint framework was found to have better recovery of the item covariances with a correlation of $0.642$ between the observed and the estimated item covariances versus $0.488$ in the separate model. The results for the network data and the questionnaire data show that the joint modeling framework has the potential to improve the model fit and the predictive fit of the data over separate models.

 \setlength{\arrayrulewidth}{.3mm}
\setlength{\tabcolsep}{10pt}
\renewcommand{\arraystretch}{.5}
\begin{table}[!ht]\caption{Dependence between the network and item responses}
 \begin{center}
\begin{tabular}{lccc} 
\toprule
  &\multicolumn{3}{c}{Standardized Function Coefficient}\\
     \cmidrule(lr){2-4} 
  School ID & Student & Policy & Satisfaction\\
 \midrule
15 & \textbf{2.231}&  -0.268&  \textbf{-1.489}\\
24 & \textbf{-0.874} & -0.139 & -0.226\\
 27 &  \textbf{-1.206 } &  0.523 &  0.320\\
32 &   \textbf{-1.191 } &  0.275 &  0.681\\
  \addlinespace
35    &\textbf{1.377} &  -0.753 &  0.292\\
44   &-0.404 &  0.643 &  \textbf{0.812} \\
56  & 0.241 &  0.546  & \textbf{1.437} \\
57   &  \textbf{0.851} &  -0.317&   0.489\\
    \addlinespace
58  &  0.824  &  0.434 &  \textbf{2.001}\\
63  &0.807 &  0.391&  \textbf{-1.475} \\
64 &  \textbf{7.979}  & -1.364& \textbf{ -7.266} \\
65 &\textbf{0.601} &   0.166 &  \textbf{0.581} \\
      \addlinespace
7  &\textbf{6.530}  &  2.332 & \textbf{-4.326}\\
8 & -0.390&  \textbf{ 0.892} &  -0.473\\
\bottomrule
\end{tabular}
\label{cca_others}
\end{center}
\end{table} 

\textbf{Summary of the Relationships.}\label{others} Out of the fourteen schools, Schools 24, 56 and 63 showed significant relationships (with the significance level of $0.05$) between the network and item responses with Wilks' $\lambda$ of $0.186$, $0.084$ and $0.120$ and p-values of $0.004$, $0.009$ and $0.005$. For all $14$ schools, we summarize the relationship between the network and item responses based on the first canonical correlations given that the first canonical correlations were estimated to be consistently large.  We focus on interpreting the dimensions of the item responses and their relationships with the advice-seeking networks.

For the dimensions of the item responses, the student dimension was found to be less clearly replicated than the policy and the satisfaction dimensions with the median congruence coefficients (between the rotated dimensions and the theoretical structure) being $0.565$, $0.885$ and $0.790$ for the student, policy and satisfaction dimensions, respectively. To assess which dimension(s) contribute to the first canonical correlations, we compared the size of the standardized function coefficients for each dimension across schools. Table \ref{cca_others} shows the standardized function coefficients for all schools; the largest coefficients were bolded for each school, and the second largest coefficients were also bolded if their sizes were comparable to the largest coefficients. Keep in mind that there is subjectivity associated with interpreting CCA results---a standardized function coefficient may seem large to one person and not large enough to the next. The results show that in $9$ out of the $14$ schools, the student dimension had the largest function coefficients; in only $1$ out of the $14$ schools, the policy dimension had the largest function coefficient; and in $8$ out of the $14$ schools, the satisfaction dimension had the largest or the second largest coefficients. Overall, the CCA results show that the perception of students and the satisfaction are related to the advice-seeking relationships in most schools, but the same cannot be said for the perceived influence over school policies.

\section{Simulation}

We conducted a simulation study to evaluate the recovery of parameters when the data are generated using the parameter estimates from school 56. We simulated $100$ data sets under the JNIRM with $N=24$ (the number of teachers in school 56) and $100$. We assessed the recovery of the model parameters based on the bias, variance and MSE over these 100 simulated datasets.

The data were generated as follows. We first generated the network dimensions as well as the latent variables of the item responses from the multivariate normal distribution with mean zero and covariance as the estimated covariance from school 56. We used the estimated parameters from school 56 (including the item intercepts, the unrotated item slopes and the network intercept), made calculations following Equations \ref{n1} and \ref{irt} without the error terms and used those calculations as the (true) expected values of the item responses and the network edge values. The errors for the network and the item responses were simulated as follows. For the network, we simulated errors following Equation \ref{n1} using the estimated within-dyad correlation from school 56 and the error variance of $1$. Binary network edges were generated based on whether the sums of the expected values and the simulated errors were larger than 0. For the item responses, we randomly sampled the errors from a normal distribution with mean $0$ and variance as the estimated error variance from school 56. The simulated item responses were the sums of the expected values and the simulated errors\footnote{The observed data for school 56 were categorical. They were treated as continuous and fitted to a linear model.}. This process was repeated $100$ times for $N=24$ and for $N=100$.

For each generated data set, the MCMC algorithm was run for $50,000$ iterations with $1,000$ burn-in iterations. We estimated the model parameters including the network intercept, within-dyad correlation, item response error variance, the item slopes and calculated the bias, variance and MSE. In addition, we assessed the recovery of each cell of the expected values of  the item responses and network edge values, which indirectly reflects the recovery of each cell of $\boldsymbol{U}\boldsymbol{V}^T$ and $\boldsymbol{\Theta}\boldsymbol{A}^T$.

In Table \ref{bias_pa}, we present the bias and the MSE of the model parameters for $N=24$ and $N =100$. 
The results show that there was little bias in all parameters when $N=100$, although overall, the bias was larger when $N=24$. Some of the MSEs at $N=24$ were above $0.01$, but most were reasonably small. The MSEs decreased significantly when $N$ was increased to 100.

\begin{figure}
\centering
  \includegraphics[scale=0.15]{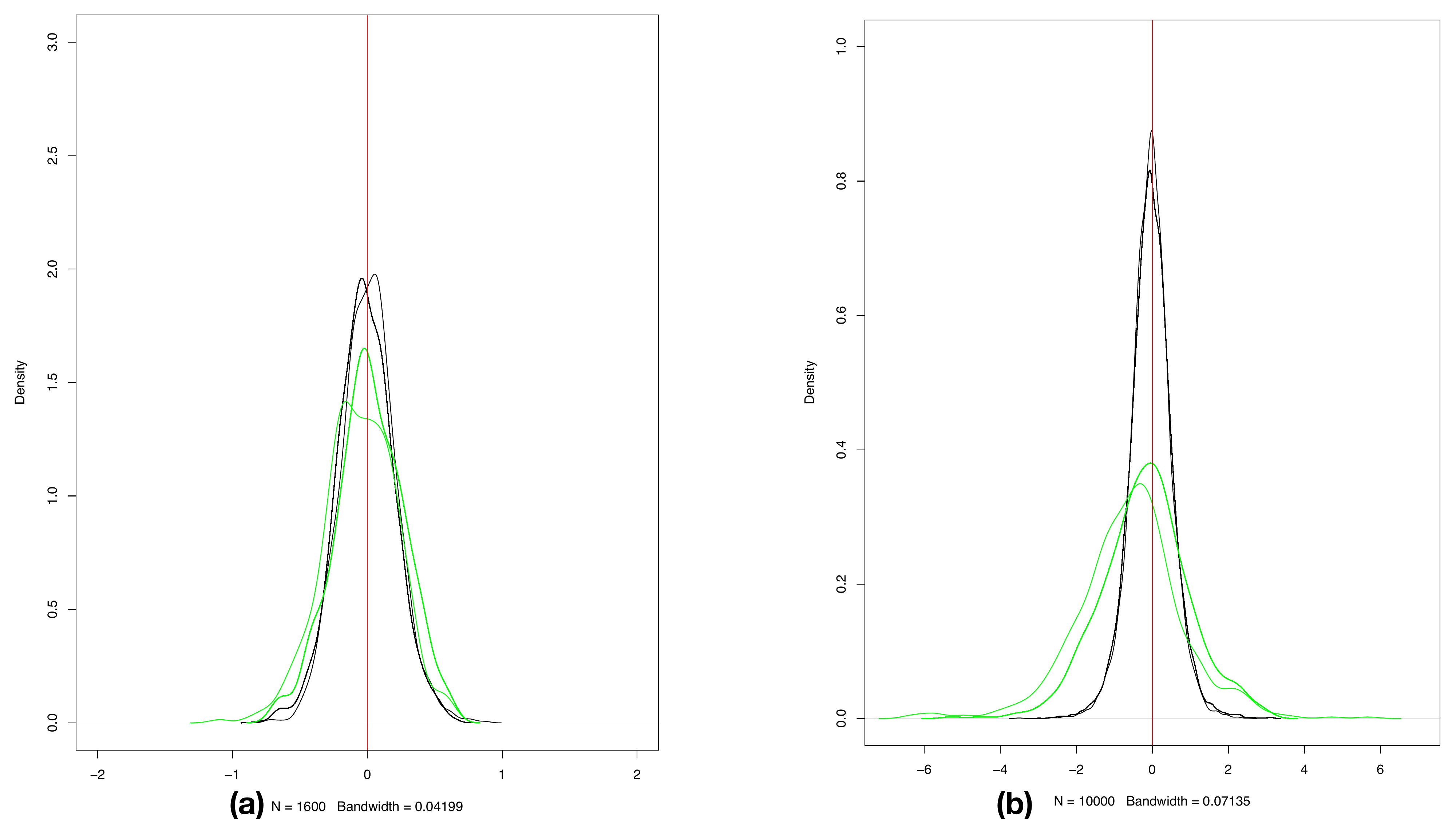}
 \caption{(a) The density of the differences between the estimated posterior means and the true expected item responses when the data were generated based on parameter estimates from school 56 for $N=24$ (green) vs $N=100$ (black). (b) The density of the differences between the estimated posterior means and the true expected network edge values when the data were generated based on parameter estimates from school 56 for $N=24$ (green) vs $N=100$ (black).  }
\label{sim_p}
\end{figure}

In Figures \ref{sim_p}, we present the density of the differences between the estimated posterior means and the true expected values for the network and the item responses when the data were generated based on parameter estimates from school 56. The results show that the model was able to recover the expected values for both the network and the item responses in both sample size conditions although the estimation was more precise when $N=100$. Both sets of results provide evidence that the JNIRM is able to reasonably recover model parameters in data situations similar to school 56.

 \setlength{\arrayrulewidth}{.3mm}
\setlength{\tabcolsep}{10pt}
\renewcommand{\arraystretch}{.2}
\begin{table}[!ht]\caption{Parameter Recovery}
 \begin{center}
\begin{tabular}{lcccc} 
\toprule
 & \multicolumn{2}{c}{Bias} &\multicolumn{2}{c}{MSE}\\
 \cmidrule(lr){2-3}  \cmidrule(lr){4-5}
 Parameter& $N=24$ & $N=100$ & $N=24$ & $N=100$\\
 \midrule
$\delta$ &   -0.170 &-0.041 &0.031 &0.002\\
$\rho$ &0.069 & 0.009& 0.005 &0.000\\
$\sigma^2_{\epsilon}$ &0.062&  0.040& 0.004& 0.002\\
  \addlinespace
$\beta_1$   &       -0.100  &0.047 & 0.011& 0.002       \\
$\beta_2$     &      -0.158 & 0.047 & 0.025 &0.002\\
$\beta_3$       &     -0.052 &-0.058  & 0.003 &0.003\\
$\beta_4$     &  -0.249 &-0.022& 0.062 &0.001\\
$\beta_5$       &    -0.011& -0.022& 0.000& 0.001\\
$\beta_6$         &-0.026 &-0.020& 0.001 &0.001\\
$\beta_7$           &  0.265 & 0.090& 0.071 &0.008\\
$\beta_8$          &  -0.031  &0.128 &0.001 &0.017\\
$\beta_9$        &-0.129  &0.049  &0.018 &0.002\\
$\beta_{10}$        &  -0.243 & 0.007& 0.059 &0.000\\
$\beta_{11}$          &0.126 &-0.026    &0.018& 0.001\\
$\beta_{12}$          &0.060& -0.021 &0.004 &0.001\\
$\beta_{13}$         &0.109 &-0.111 &0.012 &0.012\\
$\beta_{14}$      &-0.029 &-0.135 &0.001& 0.018\\
$\beta_{15}$        &   -0.230 & 0.004 & 0.053 &0.000\\
$\beta_{16}$          & 0.057 &-0.026 &0.004 &0.001\\
\bottomrule
\end{tabular}
\label{bias_pa}
\end{center}
\end{table} 

\textbf{Bias in Sparse Networks with Finite Samples.} The density of the binary network is influenced by both the intercept and the variances of the network dimensions.
A
We conducted a small simulation study to assess whether there is bias in the estimation of the intercept and the variances of the network dimensions.  

We first investigated how the density of the binary network depends on the intercept and the variances of the network dimensions. To do that, we generated data sets under the proposed network model in Equation \ref{n1} with different intercepts and different variances of the network dimensions (with no within-dyad correlation), one data set per combination of an intercept value with a variance value; and we compared the densities of the resulting binary networks. Table \ref{tabled} presents the densities of the binary networks when the intercepts are $0$, $-1$, $-2$, $-3$ and $-4$; and the variances of the network dimensions, U and V are $0.2$ versus $1$. For simplicity, we assumed the same variance for all dimensions of U and V and orthogonality among the dimensions; when the variance was $1$, it means that an identity matrix was used as the covariance for the network dimensions. In Table \ref{tabled}, both the intercept and the variances of the network dimensions influenced the density of the network. Therefore, a biased estimate for the intercept can potentially cooccur with a biased estimate of the latent variable variance while preserving the density of the network, e.g., the intercept can be overestimated while the latent variable variance is underestimated.

\setlength{\arrayrulewidth}{.3mm}
\setlength{\tabcolsep}{50pt}
\renewcommand{\arraystretch}{1.2}
\begin{table}[!ht]\caption{Density of binary networks. }
 \begin{center}
\begin{tabular}{lcccc} 
\hline
& \multicolumn{2}{c}{Variance of the Network Dimensions}\\
\cmidrule(lr){2-3}
Intercept & $0.2$ & $1$ \\           
\hline
  0  & 0.498 & 0.500 \\ 
  -1  & 0.165 & 0.247 \\ 
  -2  & 0.025& 0.096 \\ 
  -3  & 0.001 & 0.033 \\ 
   -4  & 0 & 0.012 \\ 
\hline 
\end{tabular}
\begin{tablenotes}
      \linespread{1}\small
          \item Note: The table contains the densities of the generated binary networks in vector latent space models with only the intercept and the vector product. The table contains the densities of the binary networks when the intercepts are $0$, $-1$, $-2$, $-3$ to $-4$; and the variances for the network dimensions, U and V are $0.2$ versus $1$.
    \end{tablenotes}
\label{tabled}
\end{center}
\end{table}

\setlength{\arrayrulewidth}{.3mm}
\setlength{\tabcolsep}{5pt}
\renewcommand{\arraystretch}{1.2}
\begin{table}[!ht]\caption{Parameter recovery}
 \begin{center}
\begin{tabular}{lccccc@{\hspace{1cm}}cccc} 
\toprule
&\multicolumn{5}{c}{Bias} & \multicolumn{4}{c}{Kurtosis}\\       
  \cmidrule(lr){2-6} \cmidrule(lr){7-10}
 & &\multicolumn{2}{c}{Var(U)} & \multicolumn{2}{c}{Var(V)}&\multicolumn{2}{c}{U} & \multicolumn{2}{c}{V}\\ 
 \cmidrule(lr){3-4}   \cmidrule(lr){5-6} \cmidrule(lr){7-8}  \cmidrule(lr){9-10}
 True Intercept &Intercept&D1&D2&D1&D2&D1&D2&D1&D2\\
\midrule
$N=100$&&&&&&&&&\\
0 & 0.010*&-0.033    &   -0.134      &-0.021&       -0.136  & 2.510 &       2.390   &    2.655  &      2.741\\  
-1&0.017 & -0.079    &   -0.106      &-0.083&       -0.097 & 3.095     &   2.809       &2.641      &  2.206\\
-2&0.018&-0.130       &-0.213      &-0.129    &   -0.210  &3.251        &2.996       &2.362        &2.959\\
-3&0.183&-0.213       &-0.328    &  -0.213      & -0.331 &3.583        &3.305       &2.933        &3.000\\
-4& 0.835 &-0.535      & -0.694&      -0.533      & -0.696&5.578        &4.547       &3.889        &4.412\\
  \addlinespace
  $N=1000$&&&&&&&&&\\
0 &-0.002& 0.011     &  -0.076    &   0.011     &  -0.076   & 3.001  &      2.989   &    2.709    &    3.077\\
-1    & 0.001&  0.018    &   -0.077     &  0.018   &    -0.077  & 3.135    &    2.973   &    2.716  &      3.128\\
-2   &  -0.009&  0.027    &   -0.074    &   0.027   &    -0.074  & 3.132    &    2.966     &  2.757   &     3.093\\
-3   &  0.002&  0.013    &   -0.092    &   0.013    &   -0.092   & 3.160     &   3.057   &   2.775     &   3.143\\
-4    &  0.270 &0.211     &  -0.614    &  -0.213    &   -0.600    &  3.618     &   2.372    &   3.346    &    2.457\\
\bottomrule
\end{tabular}
\label{tablebias}
\begin{tablenotes}
      \linespread{1}\small
          \item Note: Var(U) is estimated based on the covariance of the estimated U. U is estimated as the $K$ left singular vectors of the posterior mean of the vector products. The same is true for Var(V) and V. *Positive values indicate overestimation, and negative values indicate underestimation. 
    \end{tablenotes}
\end{center}
\end{table}

Second, we investigated whether there is bias in the estimation of the intercept and the variances of the network dimensions. To do that, we fitted the generated data sets with $N=100$ and $1,000$ to the proposed network model and calculated the bias of the intercepts and the variances of the network dimensions. We used the five data sets generated in conditions specified in the second column of Table \ref{tabled} with the $2$-dimensional multiplicative UV effects that have an identity covariance matrix and decreasing intercepts $(0, -1, -2, -3, \text{ and} -4)$. We fitted these data sets with $N=100$ to the network model with an intercept and a 2-dimensional multiplicative effect UV effect shown in Equation \ref{n1} and repeated the process for $N=1,000$. The biases of the intercept and the biases of the variances of the network dimensions are presented in Table \ref{tablebias}. In finite samples, as the intercept decreased, the estimated intercept became more biased in the positive direction. For $N =100$, the intercepts showed an overestimation of $0.010$, $0.017$, $0.018$, $0.183$ and $0.835$ when the true intercepts were $0, -1, -2, -3, \text{and} -4,$ respectively. In finite samples, the variances showed an underestimation problem, and the estimates became more biased as the networks became sparser (the intercept became more negative). For a zero intercept or a slightly negative intercept, the biases were small. The Kurtosis of the latent variables are also presented in Table \ref{tablebias}. The results show that when $N=100$, latent variable distributions became more heavy-tailed than the normal distribution when the networks were sparse. When $N=1,000$, the biases were negligible and only noticeable when the true intercept was $-4$.

The results show that in sparse networks with finite samples, there are a positive bias in the estimation of the intercept and negative biases in the estimation of the variances of the network dimensions. This is an issue to consider when the networks are extremely sparse with a density around $0.01$ or less in finite samples and when we need to make separate inferences for the intercept and the variances of the network dimensions. The issue disappears when we make an inference that jointly considers the intercept and the variances of the network dimensions, for example, estimating the density of the network. If the network is not sparse, there is (almost) no bias regardless of the sample size; and when the sample size is sufficiently large, the intercepts and the variances of the network dimensions can be correctly recovered. In our applications, the densities of the networks are always larger than $0.01$. Besides, we focus on making inferences about the relationship between the network and the item responses, and the separate inferences about the variances of the network dimensions and the intercept are not the focus of this project.

\section{Discussions and limitations}

We have introduced a novel joint framework for modeling the dependence between a network and item responses, or more broadly speaking, the node attributes. Many authors have proposed methods for modeling the relationship between the network and the node attributes. However, most of these existing methods only allow inferences about the relationship as one-directional, either as the effect of the attributes on the network or as the effect of the network on the attributes, and without a data generation model for both. Our method allows inferences about the co-variation or the correlational relationship between the network and attributes. More specifically, we model the relationship between the network and item responses at the latent level on both sides; the latent network dimensions and the latent variables of the item responses inform each other, which offers an improvement in the estimation quality when the network and the item responses are related. 

Our analysis shows promising results for interpreting the dependence between the network and item responses at the latent construct level on both sides. To resolve the rotational indeterminacy of the axes in the latent space, we use the first $K$ left and right singular vectors of the posterior means of the vector products as the estimates of the latent network dimensions. The same practice is also applied for the item responses.  In this way, we are able to obtain estimates of the latent network dimensions and the latent variables of the item responses following the convergence of the vector products and use CCA to study their relationship. The downside of this approach is that we cannot directly provide uncertainty information about the covariance matrix between the two sets of latent variables based on the MCMC chain, and the potential measurement error associated with estimating the latent variables was not accounted for by the CCA. A possible solution is to impose further constraints in the model, for example, fixing the dimensions of the network and the dimensions of the item responses before or within the model estimation process. Fixing the dimensions before the estimation would require a confirmatory framework for both the network and the item responses. A downside to the confirmatory approach is that it is difficult to know what network dimensions to expect ahead of time. Further theoretical developments about the latent dimensions of a network are needed to identify and attribute meanings to those dimensions.

The limitations notwithstanding, the proposed methodology has broad application potential. It is of interest in many disciplines to test and model the relationship between a network and node attributes. We name a few here. In neuroscience, it is of interest to test and model the relationship between the connections among brain regions and the attributes of those regions such as the volume and the thickness of the regions.  In economics and computer science, it is of interest to jointly model relational data, e.g., likes and followers in social media and other user information such as personalities, health outcomes, online behavior choices, etc. We can make predictions about a user's future consuming behaviors based on the user's network information and current consuming behaviors. It is also of interest to model consumer information with their geographic networks.

\section*{Acknowledgement} 
The authors would like to thank Prof. Paul De Boeck for his helpful discussions, mentorship, and encouragement. We also thank Prof. Jim Spillane and the Distributed Leadership Studies at Northwestern University, including the NebraskaMATH study, for kindly providing us with the dataset used for this study. SP's research was partially supported by National Science Foundation grant DMS-1830547. 

\section{Keywords}
Latent Space Model, Social Network Analysis, High-dimensional Analysis

\section{Competing Interest}
None.

\section{Bibliography}

\printbibliography[heading=none]

@incollection{samejima2016graded,
  title={Graded response models},
  author={Samejima, Fumiko},
  booktitle={Handbook of item response theory},
  pages={95--107},
  year={2016},
  publisher={Chapman and Hall/CRC}
}

@article{muraki1990fitting,
  title={Fitting a polytomous item response model to Likert-type data},
  author={Muraki, Eiji},
  journal={Applied Psychological Measurement},
  volume={14},
  number={1},
  pages={59--71},
  year={1990},
  publisher={Sage Publications Sage CA: Thousand Oaks, CA}
}

@article{pustejovsky2009question,
  title={Question-order effects in social network name generators},
  author={Pustejovsky, James E and Spillane, James P},
  journal={Social networks},
  volume={31},
  number={4},
  pages={221--229},
  year={2009},
  publisher={Elsevier}
}

@article{pitts2009using,
  title={Using social network methods to study school leadership},
  author={Pitts, Virginia M and Spillane, James P},
  journal={International Journal of Research \& Method in Education},
  volume={32},
  number={2},
  pages={185--207},
  year={2009},
  publisher={Taylor \& Francis}
}

@article{wang2023joint,
  title={Joint latent space model for social networks with multivariate attributes},
  author={Wang, Selena and Paul, Subhadeep and De Boeck, Paul},
  journal={Psychometrika},
  volume={88},
  number={4},
  pages={1197--1227},
  year={2023},
  publisher={Springer}
}

@article{broda2021using,
  title={Using social network analysis in applied psychological research: A tutorial.},
  author={Broda, Michael D and Granger, Kristen and Chow, Jason and Ross, Erica},
  journal={Psychological Methods},
  year={2021},
  publisher={American Psychological Association}
}

@incollection{bonsignore2011power,
  title={The power of social networking for professional development},
  author={Bonsignore, Elizabeth and Hansen, Derek and Galyardt, April and Aleahmad, Turadg and Hargadon, Steve},
  booktitle={Breakthrough teaching and learning},
  pages={25--52},
  year={2011},
  publisher={Springer}
}

@article{wang2022resilience,
  title={Resilience to stress in bipartite networks: Application to the Islamic State recruitment network},
  author={Wang, Selena and Edgerton, Jared},
  journal={Journal of Complex Networks},
  volume={10},
  number={4},
  pages={cnac017},
  year={2022},
  publisher={Oxford University Press}
}

@techreport{hoff_2015_b,
  title = {Dyadic data analysis with amen},
  author = {Hoff, P.D.},
  institution = {Department of Statistics, University of Washington},
  year = {2015},
  number = {638},
  eprint = {arXiv:1506.08237},
  tmac = {methods,computation,application}
}

@article{jin2015fast,
  title={Fast community detection by SCORE},
  author={Jin, Jiashun},
  journal={The Annals of Statistics},
  volume={43},
  number={1},
  pages={57--89},
  year={2015},
  publisher={Institute of Mathematical Statistics}
}

@article{zhao2012consistency,
  title={Consistency of community detection in networks under degree-corrected stochastic block models},
  author={Zhao, Yunpeng and Levina, Elizaveta and Zhu, Ji},
  journal={The Annals of Statistics},
  volume={40},
  number={4},
  pages={2266--2292},
  year={2012},
  publisher={Institute of Mathematical Statistics}
}

@article{karrer2011stochastic,
  title={Stochastic blockmodels and community structure in networks},
  author={Karrer, Brian and Newman, Mark EJ},
  journal={Physical review E},
  volume={83},
  number={1},
  pages={016107},
  year={2011},
  publisher={APS}
}

@article{browne2008cefa,
  title={CEFA: Comprehensive exploratory factor analysis, Version 2.00 [Computer software and manual]},
  author={Browne, MW and Cudeck, R and Tateneni, K and Mels, G},
  journal={Available at online: http://faculty. psy. ohio-state. edu/browne},
  year={2008}
}

@article{sweet2020latent,
  title={A Latent Space Network Model for Social Influence},
  author={Sweet, Tracy and Adhikari, Samrachana},
  journal={Psychometrika},
  pages={1--24},
  year={2020},
  publisher={Springer}
}

@article{hoff2002latent,
  title={Latent space approaches to social network analysis},
  author={Hoff, Peter D and Raftery, Adrian E and Handcock, Mark S},
  journal={Journal of the american Statistical association},
  volume={97},
  number={460},
  pages={1090--1098},
  year={2002},
  publisher={Taylor \& Francis}
}

@article{krivitsky2009representing,
  title={Representing degree distributions, clustering, and homophily in social networks with latent cluster random effects models},
  author={Krivitsky, Pavel N and Handcock, Mark S and Raftery, Adrian E and Hoff, Peter D},
  journal={Social networks},
  volume={31},
  number={3},
  pages={204--213},
  year={2009},
  publisher={Elsevier}
}

@article{mcpherson2001birds,
  title={Birds of a feather: Homophily in social networks},
  author={McPherson, Miller and Smith-Lovin, Lynn and Cook, James M},
  journal={Annual review of sociology},
  volume={27},
  number={1},
  pages={415--444},
  year={2001},
  publisher={Annual Reviews 4139 El Camino Way, PO Box 10139, Palo Alto, CA 94303-0139, USA}
}

@article{epskamp2017generalized,
  title={Generalized network pschometrics: Combining network and latent variable models},
  author={Epskamp, Sacha and Rhemtulla, Mijke and Borsboom, Denny},
  journal={Psychometrika},
  volume={82},
  number={4},
  pages={904--927},
  year={2017},
  publisher={Springer}
}

@article{hoff2005bilinear,
  title={Bilinear mixed-effects models for dyadic data},
  author={Hoff, Peter D},
  journal={Journal of the american Statistical association},
  volume={100},
  number={469},
  pages={286--295},
  year={2005},
  publisher={Taylor \& Francis}
}

@article{minhas2019inferential,
  title={Inferential Approaches for Network Analysis: AMEN for Latent Factor Models},
  author={Minhas, Shahryar and Hoff, Peter D and Ward, Michael D},
  journal={Political Analysis},
  volume={27},
  number={2},
  pages={208--222},
  year={2019},
  publisher={Cambridge University Press}
}

@incollection{de2004framework,
  title={A framework for item response models},
  author={De Boeck, Paul and Wilson, Mark},
  booktitle={Explanatory item response models},
  pages={3--41},
  year={2004},
  publisher={Springer}
}

@inproceedings{zhang2020flexible,
  title={A flexible latent space model for multilayer networks},
  author={Zhang, Xuefei and Xue, Songkai and Zhu, Ji},
  booktitle={International Conference on Machine Learning},
  pages={11288--11297},
  year={2020},
  organization={PMLR}
}

@book{jackson2008social,
  title={Social and economic networks},
  author={Jackson, Matthew O and others},
  volume={3},
  year={2008},
  publisher={Princeton University Press Princeton}
}

@article{jeong2000large,
  title={The large-scale organization of metabolic networks},
  author={Jeong, Hawoong and Tombor, B{\'a}lint and Albert, R{\'e}ka and Oltvai, Zoltan N and Barab{\'a}si, A-L},
  journal={Nature},
  volume={407},
  number={6804},
  pages={651--654},
  year={2000},
  publisher={Nature Publishing Group}
}

@article{bullmore09,
  title={Complex brain networks: graph theoretical analysis of structural and functional systems},
  author={Bullmore, Ed and Sporns, Olaf},
  journal={Nature Reviews Neuroscience},
  volume={10},
  number={3},
  pages={186--198},
  year={2009},
  publisher={Nature Publishing Group}
}

@article{rubinov10,
  title={Complex network measures of brain connectivity: uses and interpretations},
  author={Rubinov, Mikail and Sporns, Olaf},
  journal={Neuroimage},
  volume={52},
  number={3},
  pages={1059--1069},
  year={2010},
  publisher={Elsevier}
}

@article{hoff2021additive,
  title={Additive and multiplicative effects network models},
  author={Hoff, Peter},
  journal={Statistical Science},
  volume={36},
  number={1},
  pages={34--50},
  year={2021},
  publisher={Institute of Mathematical Statistics}
}

@article{bastiampillai2013depression,
  title={Is depression contagious? The importance of social networks and the implications of contagion theory},
  author={Bastiampillai, Tarun and Allison, Stephen and Chan, Sherry},
  journal={Australian \& New Zealand Journal of Psychiatry},
  volume={47},
  number={4},
  pages={299--303},
  year={2013},
  publisher={Sage Publications Sage UK: London, England}
}

@article{wuchty2014controllability,
  title={Controllability in protein interaction networks},
  author={Wuchty, Stefan},
  journal={Proceedings of the National Academy of Sciences},
  volume={111},
  number={19},
  pages={7156--7160},
  year={2014},
  publisher={National Acad Sciences}
}

@article{zhang2019scale,
  title={Scale-free resilience of real traffic jams},
  author={Zhang, Limiao and Zeng, Guanwen and Li, Daqing and Huang, Hai-Jun and Stanley, H Eugene and Havlin, Shlomo},
  journal={Proceedings of the National Academy of Sciences},
  volume={116},
  number={18},
  pages={8673--8678},
  year={2019},
  publisher={National Acad Sciences}
}

@article{baggio2016multiplex,
  title={Multiplex social ecological network analysis reveals how social changes affect community robustness more than resource depletion},
  author={Baggio, Jacopo A and BurnSilver, Shauna B and Arenas, Alex and Magdanz, James S and Kofinas, Gary P and De Domenico, Manlio},
  journal={Proceedings of the National Academy of Sciences},
  volume={113},
  number={48},
  pages={13708--13713},
  year={2016},
  publisher={National Acad Sciences}
}

@article{janssen2006toward,
  title={Toward a network perspective of the study of resilience in social-ecological systems},
  author={Janssen, Marco A and Bodin, {\"O}rjan and Anderies, John M and Elmqvist, Thomas and Ernstson, Henrik and McAllister, Ryan RJ and Olsson, Per and Ryan, Paul},
  journal={Ecology and Society},
  volume={11},
  number={1},
  year={2006},
  publisher={JSTOR}
}

@article{cohen2000resilience,
  title={Resilience of the internet to random breakdowns},
  author={Cohen, Reuven and Erez, Keren and Ben-Avraham, Daniel and Havlin, Shlomo},
  journal={Physical review letters},
  volume={85},
  number={21},
  pages={4626},
  year={2000},
  publisher={APS}
}

@article{mele2019spectral,
  title={Spectral inference for large Stochastic Blockmodels with nodal covariates},
  author={Mele, Angelo and Hao, Lingxin and Cape, Joshua and Priebe, Carey E},
  journal={arXiv preprint arXiv:1908.06438},
  year={2019}
}

@book{lusher2013exponential,
  title={Exponential random graph models for social networks: Theory, methods, and applications},
  author={Lusher, Dean and Koskinen, Johan and Robins, Garry},
  year={2013},
  publisher={Cambridge University Press}
}

@article{sweet2015incorporating,
  title={Incorporating covariates into stochastic blockmodels},
  author={Sweet, Tracy M},
  journal={Journal of Educational and Behavioral Statistics},
  volume={40},
  number={6},
  pages={635--664},
  year={2015},
  publisher={SAGE Publications Sage CA: Los Angeles, CA}
}

@article{hoff2015dyadic,
  title={Dyadic data analysis with amen},
  author={Hoff, Peter D},
  journal={arXiv preprint arXiv:1506.08237},
  year={2015}
}

@article{fosdick2015testing,
  title={Testing and modeling dependencies between a network and nodal attributes},
  author={Fosdick, Bailey K and Hoff, Peter D},
  journal={Journal of the American Statistical Association},
  volume={110},
  number={511},
  pages={1047--1056},
  year={2015},
  publisher={Taylor \& Francis}
}

@article{penuel2013organization,
  title={The organization as a filter of institutional diffusion},
  author={Penuel, William and Frank, Kenneth and Sun, Min and Kim, Chong and Singleton, Corinne},
  journal={Teachers college record},
  volume={115},
  number={1},
  pages={1--33},
  year={2013}
}

@article{weinbaum2008going,
  title={Going with the flow: Communication and reform in high schools},
  author={Weinbaum, Elliot H and Cole, Russell P and Weiss, Michael J and Supovitz, Jonathan A},
  journal={The implementation gap: Understanding reform in high schools},
  pages={68--102},
  year={2008}
}

@article{moolenaar2010occupying,
  title={Occupying the principal position: Examining relationships between transformational leadership, social network position, and schools’ innovative climate},
  author={Moolenaar, Nienke M and Daly, Alan J and Sleegers, Peter JC},
  journal={Educational administration quarterly},
  volume={46},
  number={5},
  pages={623--670},
  year={2010},
  publisher={Sage Publications Sage CA: Los Angeles, CA}
}

@article{revelle2015package,
  title={Package ‘psych’},
  author={Revelle, William and Revelle, Maintainer William},
  journal={The comprehensive R archive network},
  volume={337},
  pages={338},
  year={2015}
}

@techreport{anderson1962introduction,
  title={An introduction to multivariate statistical analysis},
  author={Anderson, Theodore Wilbur},
  year={1962},
  institution={Wiley New York}
}

@article{wang2021recent,
  title={Recent Integrations of Latent Variable Network Modeling With Psychometric Models.},
  author={Wang, Selena},
  journal={Frontiers in psychology},
  volume={12},
  pages={773289--773289},
  year={2021}
}

@article{palestro2018tutorial,
  title={A tutorial on joint models of neural and behavioral measures of cognition},
  author={Palestro, James J and Bahg, Giwon and Sederberg, Per B and Lu, Zhong-Lin and Steyvers, Mark and Turner, Brandon M},
  journal={Journal of Mathematical Psychology},
  volume={84},
  pages={20--48},
  year={2018},
  publisher={Elsevier}
}

@article{marden2015multivariate,
  title={Multivariate statistics},
  author={Marden, John I},
  journal={Urbana-Champaign: Department of Statistics, University of Illinois},
  year={2015}
}

@article{sherry2005conducting,
  title={Conducting and interpreting canonical correlation analysis in personality research: A user-friendly primer},
  author={Sherry, Alissa and Henson, Robin K},
  journal={Journal of personality assessment},
  volume={84},
  number={1},
  pages={37--48},
  year={2005},
  publisher={Taylor \& Francis}
}

@book{thompson1984canonical,
  title={Canonical correlation analysis: Uses and interpretation},
  author={Thompson, Bruce},
  number={47},
  year={1984},
  publisher={Sage}
}

@article{paul2022causal,
  title={Network Influence with Latent Homophily and Measurement Error},
  author={Paul, Subhadeep and Nath, Shanjukta and Warren, Keith},
  journal={arXiv preprint arXiv:2203.14223},
  year={2022}
}

@article{huang2022mutually,
  title={A Mutually Exciting Latent Space Hawkes Process Model for Continuous-time Networks},
  author={Huang, Zhipeng and Soliman, Hadeel and Paul, Subhadeep and Xu, Kevin S},
  journal={Conference on Uncertainty in Artificial Intelligence (UAI)},
  year={2022}
}

@article{paul2022null,
  title={Null models and community detection in multi-layer networks},
  author={Paul, Subhadeep and Chen, Yuguo},
  journal={Sankhya A},
  volume={84},
  number={1},
  pages={163--217},
  year={2022},
  publisher={Springer}
}
\end{document}